\documentclass{aa} 

\usepackage{graphicx}
\usepackage{xcolor}
\usepackage{amsmath,amssymb}
\usepackage{txfonts}
\usepackage{natbib}
\bibpunct{(}{)}{;}{a}{}{,}

\begin{document}

   \title{Tidally perturbed g-mode pulsations\\in a sample of close eclipsing binaries\thanks{The reduced and analysed datasets are also available in electronic form at \href{https://doi.org/10.5281/zenodo.7555868}{https://doi.org/10.5281/zenodo.7555868}.}}
   \titlerunning{Tidally perturbed g-mode pulsations}

   \author{T. Van Reeth \inst{1,2}
           \and
           C. Johnston\inst{3,1}
           \and
           J. Southworth \inst{4}
           \and
           J. Fuller \inst{5,2}
           \and
           D. M. Bowman \inst{1,2}
           \and
           L. Poniatowski \inst{1}
           \and
           J. Van Beeck \inst{1,2,5}
          }

   \institute{Institute of Astronomy, KU Leuven, Celestijnenlaan 200D, B-3001 Leuven, Belgium\\
              \email{timothy.vanreeth@kuleuven.be}
             \and
             Kavli Institute for Theoretical Physics, University of California, Santa Barbara, CA 93106, USA
             \and Department of Astrophysics, IMAPP, Radboud University Nijmegen, P. O. Box 9010, 6500 GL Nijmegen, the Netherlands
             \and Astrophysics Group, Keele University, Staffordshire ST5 5BG, UK
             \and
             TAPIR, Mailcode 350-17, California Institute of Technology, Pasadena, CA 91125, USA
             }

   \date{Received; accepted}

 \abstract{Thanks to the high-precision photometry from space missions such as {\em Kepler} and TESS, tidal perturbations and tilting of pulsations have been detected in more than a dozen binary systems. However, only two of these were gravity-mode (g-mode) pulsators.}{We aim to detect tidally perturbed g~modes in additional binary systems and characterise them observationally.}{We perform a custom data reduction of the available {\em Kepler} and TESS photometry of a well-studied, published sample of 35 binary systems with $\gamma$\,Doradus ($\gamma$\,Dor) pulsators. For each target, we model the binary signal using a sum of 100 sine waves, with frequencies at orbital harmonics, and measure significant pulsation frequencies in an iterative prewhitening analysis of the residual light curve. Pulsations are labelled as tidally perturbed g~modes if they are part of both period-spacing patterns and multiplets spaced by integer multiples of the orbital frequency. After visual inspection and confirmation, the properties of these targets and g~modes are characterised.}{We detect tidally perturbed g-mode pulsations for five short-period binaries that are circularised and (almost) synchronously rotating: KIC\,3228863, KIC\,3341457, KIC\,4947528, KIC\,9108579, and KIC\,12785282. Tidally perturbed g~modes that occur within the same star and have the same mode identification ($k$,$m$), are found to have near-identical relative amplitude and phase modulations, which are within their respective $1-\sigma$ uncertainties also identical for the {\em Kepler} and TESS photometric passbands. By contrast, pulsations with different mode identification ($k$,$m$) are found to exhibit different modulations. Moreover, the observed amplitude and phase modulations are correlated, indicating that the binary tides primarily distort the g-mode amplitudes on the stellar surface. The phase modulations are then primarily a geometric effect of the integration of the stellar flux over the visible stellar surface. All selected binaries also exhibit signal that resembles rotational modulation in the Fourier domain. In the case of KIC\,3228863, this is caused by the presence of the known tertiary component, and for the other systems we hypothesise that it is caused by temperature variations on the stellar surface. Alternatively, the signal can be overstable convective modes in the stellar core or belong to the non-pulsating companion.}{While g-mode pulsation periods are known to be a direct probe of the deep interior stellar structure, the binary tides which cause the pulsation modulations are dominant in the outer stellar layers. Hence, in future tidally perturbed g~modes may allow us to do core-to-surface asteroseismic modelling of tidally distorted stars.}


   \maketitle

\section{Introduction}
\label{sec:intro}
Thanks to the near-continuous time series of high-precision photometry from space missions such as CoRoT \citep{Auvergne2009}, {\em Kepler} \citep{Koch2010}, and TESS \citep{Ricker2015}, there has been substantial progress in the field of observational asteroseismology, the study of the interior stellar structure via the analysis of stellar pulsations \citep[e.g.][]{Aerts2010}. In particular, there has been major progress in the study of gravity-mode (g-mode) pulsations, which have buoyancy as the dominant restoring force and are most sensitive to the near-core stellar properties for main-sequence stars. Here, g~modes are present in $\gamma$\,Doradus ($\gamma$\,Dor; $1.4\,M_\odot \lesssim M \lesssim 2.0\,M_\odot$) stars \citep{Kaye1999} and slowly pulsating B type (SPB; $3\,M_\odot \lesssim 8\,M_\odot$) stars \citep{Waelkens1991}. Because of their low amplitudes (usually $\lesssim 10\,$mmag) and dense pulsation spectra with typical periods between 0.3 and 3\,d, the advent of space missions gave a large boost to the detection and observational analysis of g-mode pulsators \citep[e.g.][]{Tkachenko2013,VanReeth2015,Li2020,Garcia2022,Skarka2022}.

In the asymptotic regime, where the g-mode pulsation frequency $\omega = 2\pi\nu$ is much smaller than the Brunt-V\"ais\"al\"a frequency $N$, the periods of g~modes with consecutive radial order $n$ and identical mode geometry follow regular patterns, which facilitates the analysis and interpretation of the observations. In a non-rotating, non-magnetic, chemically homogeneous star with a convective core and radiative envelope, g~modes with consecutive, high radial orders $n$ and mode identification $(k,m)$ are equidistantly spaced in period \citep{Shibahashi1979,Tassoul1980}. Here, in a non-rotating star, the meridional degree $k = l - |m|$, where $l$ and $m$ indicate the spherical degree and azimuthal order, respectively. When chemical gradients are present inside a star, g~modes can be trapped and are no longer equidistantly spaced in period, but exhibit deviations which scale with the steepness of the gradients \citep{Miglio2008a}. In addition, stellar rotation causes the pulsation frequencies to shift. As a result, the observed period spacings decrease with increasing radial order $n$ for prograde (with $m > 0$) and zonal modes (with $m = 0$), while for retrograde modes ($m < 0$) they mostly increase \citep[e.g.][]{Bouabid2013}. Using the framework of the traditional approximation of rotation \citep[TAR;][]{Eckart1960,Lee1987,Bildsten1996,Mathis2009}, this characteristic has been exploited in multiple studies \citep[e.g.][]{VanReeth2016,Ouazzani2017,Christophe2018,Li2020,Takata2020a,Pedersen2021} to measure the near-core rotation rate $\nu_{\rm rot,nc}$ and the buoyancy travel time 
\begin{equation}
    \Pi_0 = 2\pi^2\left(\int_{r_1}^{r_2}\frac{N(r)}{r}\mathrm{d}r\right)^{-1}
\end{equation} 
of stars with g-mode pulsations, where $r_1$ and $r_2$ indicate the radial boundaries of the g-mode pulsation cavity. Furthermore, in moderate- to fast-rotating stars, g-mode pulsations can occur in the sub-inertial regime, where the pulsation frequency in the co-rotating frame $\nu_{\rm co} < 2\nu_{\rm rot}$. In this case, the Coriolis force confines the g-mode pulsations to a band around the stellar equator, which is more narrow when the star is rotating faster, and the pulsations are called gravito-inertial modes. In addition, the Coriolis force contributes to the restoring force of pulsations with $k \neq 0$ or $m \leq 0$. Pulsations which have the Coriolis force as the dominant restoring force are indicated with $k < 0$ and are called inertial modes. Among these, global retrograde inertial modes or r~modes are the most often observed and well-studied \citep[e.g.][]{Saio2018,Takata2020b}.

Based on these theoretical frameworks, a lot of effort has been done to model such observed g-mode pulsators, improving our understanding of the deep interior structure of these stars and their evolution \citep[e.g.][]{Aerts2021}. Recent results of these studies include the measurement of opacity and pulsation mode excitation \citep[e.g.][]{Szewczuk2018,Szewczuk2021,Szewczuk2022}, convective core masses and sizes \citep[e.g.][]{Mombarg2021}, mixing efficiency in the radiative envelopes \citep[e.g.][]{Pedersen2021}, inference of interior magnetic field strengths \citep{Buysschaert2018,Lecoanet2021}, and constraints on the convective boundary stiffness \citep{Aerts2021b}. In turn, these successes are fuelling further efforts to improve upon the existing theoretical frameworks, such as refining the near-core \citep{Michielsen2019,Bowman2021} and extra mixing processes in the radiative envelope \citep[e.g.][]{Rogers2015,Ratnasingam2020,Mombarg2022}, calculating the coupling between inertial- and gravity-mode pulsations \citep{Ouazzani2020,Saio2021,Tokuno2022}, and expanding the TAR to include magnetic fields \citep{Prat2019,Prat2020,VanBeeck2020,Dhouib2022} and the centrifugal acceleration \citep{Henneco2021,Dhouib2021a,Dhouib2021b}.

The diagnostic power of asteroseismology increases when it is combined with complementary constraints from other analysis methods, such as the study of binary stars. In eclipsing binaries with time series of spectroscopic observations \citep{Southworth2020_bintable} or light travel time variations \citep{Hey2020}, the orbital solutions allow us to precisely determine the stellar radii and masses. Hence, in the case of binaries with sufficiently wide orbits, where the effects of tides and mass transfer on the stellar evolution can be ignored, this breaks the degeneracy between some of the model parameters that are considered in the asteroseismic modelling \citep[e.g.][]{Johnston2019, Murphy2021, Sekaran2021}. Meanwhile, in pulsating close binaries that have undergone mass transfer, asteroseismology can be used to complement the information from the binarity and help constrain the stellar properties \citep[e.g.][]{Guo2019}. Stellar pulsations can also be excited by tides \citep[e.g.][]{Fuller2017,Hambleton2018,Cheng2020} or couple to them \citep[e.g.][]{Burkart2012,Burkart2014,Weinberg2013,Li2020_bin,Guo2022}, which can then result in additional angular momentum transport \citep[e.g.][]{Fuller2021}. However, such processes are not yet well understood \citep[see][for a recent review]{Guo2021}.

Tidal perturbations of pulsations \citep[e.g.][]{Hambleton2013,Balona2018,Bowman2019,Southworth2020,Southworth2021,Steindl2021,Johnston2022}, which occur in close binaries that are circularised and (quasi-)synchronously rotating \citep{Guo2021}, are a prime example of such lesser understood interactions. Here, strong tides modify the mode cavities inside the pulsating star, which in turn results in perturbed pulsations. A subset of these, called the tidally tilted pulsators, have received much attention in the literature lately \citep[e.g.][]{Handler2020,Kurtz2020,Rappaport2021, LeeWoo2021,Alicavus2022,Jayaraman2022}. For tidally tilted pulsators, the tidal influence on the pulsations consists of three main effects: {\em (i)} tidal trapping, whereby the pulsation is confined to a specific part of the star, {\em (ii)} tidal alignment, whereby the pulsation is tilted, and has the tidal axis as its symmetry axis rather than the stellar rotation axis, and {\em (iii)} tidal amplification, whereby the pulsation mode amplitude is larger on part of the stellar surface \citep{Fuller2020}. 

Recently, tidal perturbations of g-mode pulsations were observed by \citet{Jerzykiewicz2020} and \citet{VanReeth2022a} for the SPB~star in $\pi^5$\,Ori and the $\gamma$\,Dor star in V456\,Cyg, respectively. Since tides are dominant in the outer stellar layers and g-modes are most sensitive to the near-core stellar properties, these were unexpected observations. In this paper we reanalyse a sample of 35 ellipsoidal and eclipsing binaries with $\gamma$\,Dor-type pulsators observed by the {\em Kepler} mission, and g-mode patterns detected and analysed by \citet{Li2020_bin}, with the aim of detecting and characterising new close binaries that exhibit tidally perturbed g-mode pulsations. Since all of these targets have up to 4~yr of high-precision {\em Kepler} photometry, this allows us to study the phenomenon in much more detail than for either $\pi^5$\,Ori or V456\,Cyg. Below we describe our data collection and reduction (Sect.\,\ref{sec:obs}), as well as the methods used to analyse the binarity, g~modes, tidal perturbations and rotational-modulation-like variability (Sect.\,\ref{sec:methods}), before discussing the five most promising targets (Sect.\,\ref{sec:discussion}). Finally, we characterise the tidal perturbations of g-mode pulsations (Sect.\,\ref{sec:characterisation}) and the rotational-modulation-like variability (Sect.\,\ref{sec:obs-rotmod}) in close binaries with a $\gamma$\,Dor component, and give our conclusions in Sect.\,\ref{sec:conclusions}.

\section{Observations}
\label{sec:obs}
\subsection{{\em Kepler} photometry}
\label{subsec:Kepler}
To ensure that the low-frequency instrumental variability in the analysed light curves is minimal and the measured g-mode variability is most precise, we performed a custom data reduction of the {\em Kepler} long-cadence target pixel files for the 35 $\gamma$\,Dor binaries studied by \citet{Li2020_bin}, based on the data reduction process previously described by \citet{Papics2013} and \citet{Tkachenko2013}. First, we collected the target pixel files from DR~25 at the Mikulski Archive for Space Telescopes (MAST\footnote{\href{https://archive.stsci.edu/}{https://archive.stsci.edu/}}). The light curves were then extracted from these pixel data using custom binary aperture masks, which were defined by evaluating the 85$^{\rm th}$ percentile of the measured flux within each pixel to avoid outliers. Pixels were included in the aperture mask if the evaluated electron count exceeded a threshold of $100\,\rm e^-\,s^{-1}$ and did not form secondary local maxima within the mask. The light curves were then constructed by summing over the measured electron count rates within the aperture mask for each time stamp. Next, the data were converted to mmag, and detrended by fitting and subtracting a $2^{\rm nd}$-order polynomial per quarter. To minimise the risk of overfitting the data and introducing a bias in this detrending step, it was done for all quarters simultaneously, whereby the different polynomial functions were fitted and optimised simultaneously with a sum of ten sine waves, which represented the dominant intrinsic variability of the studied target. For most of our targets, these sine waves had frequencies at integer multiples of the binary orbital frequency. Outliers were excluded from the light curves using $5-\sigma$~clipping, and finally the quality of the reduced light curves was evaluated using visual inspection. Similar custom light-curve extraction approaches have been shown to provide the most robust asteroseismic results \citep{Bowman2021}.

\subsection{TESS photometry}
\label{subsec:TESS}
To complement the {\em Kepler} photometry of the most interesting stars in the sample, we have also extracted light curves from $20\times20$-pixel cutouts from the TESS full frame images (FFI) for these targets, obtained with \texttt{TessCut} \citep{Brasseur2019}. Because the TESS photometric passband (600-1000\,nm) is relatively redder than the {\em Kepler} passband (420-900\,nm), the TESS light curves provide complementary information. However, they do not reach the same quality level. The TESS~data cover shorter time spans and consist of separate 27-d sectors rather than years-long near-continuous observations. In addition, TESS was developed to observe brighter stars; the size of the individual pixels on the CCDs is much larger (21 arcseconds compared to 3.98 arcseconds for {\em Kepler}) making the observations more susceptible to contamination from nearby stars. And finally, there is a strong background flux in the TESS data, caused by sunlight scattered by the Earth. 

Hence, additional steps were needed for the data reduction of the TESS observations. First, the background flux was estimated by calculating the 5$^{\rm th}$-percentile $\rm e^{-}\,s^{-1}$-count of the pixels in the selected CCD cutout, and subtracted from the measured $\rm e^{-}\,s^{-1}$-count in each of the pixels. This allowed us to avoid significant stellar contamination and measurement outliers. Second, the pixels of the aperture mask were selected using a $3-\sigma$ cutoff above the median flux count as the threshold, and secondary maxima within the mask were again excluded. Then the remainder of the TESS data reduction was almost identical to that of the {\em Kepler} data. The light curves were extracted by summing over the flux counts of the pixels within the aperture mask, converted to mmag~units, and detrended by subtracting a $1^{\rm st}$-order polynomial fit per orbital cycle of the TESS satellite. To avoid overfitting the detrending curve, the polynomial functions were again fitted and optimised simultaneously with a sum of ten sine waves, with frequencies equal to harmonics of the binary orbital frequency.

\begin{figure*}
    \centering
    \includegraphics[width=\textwidth]{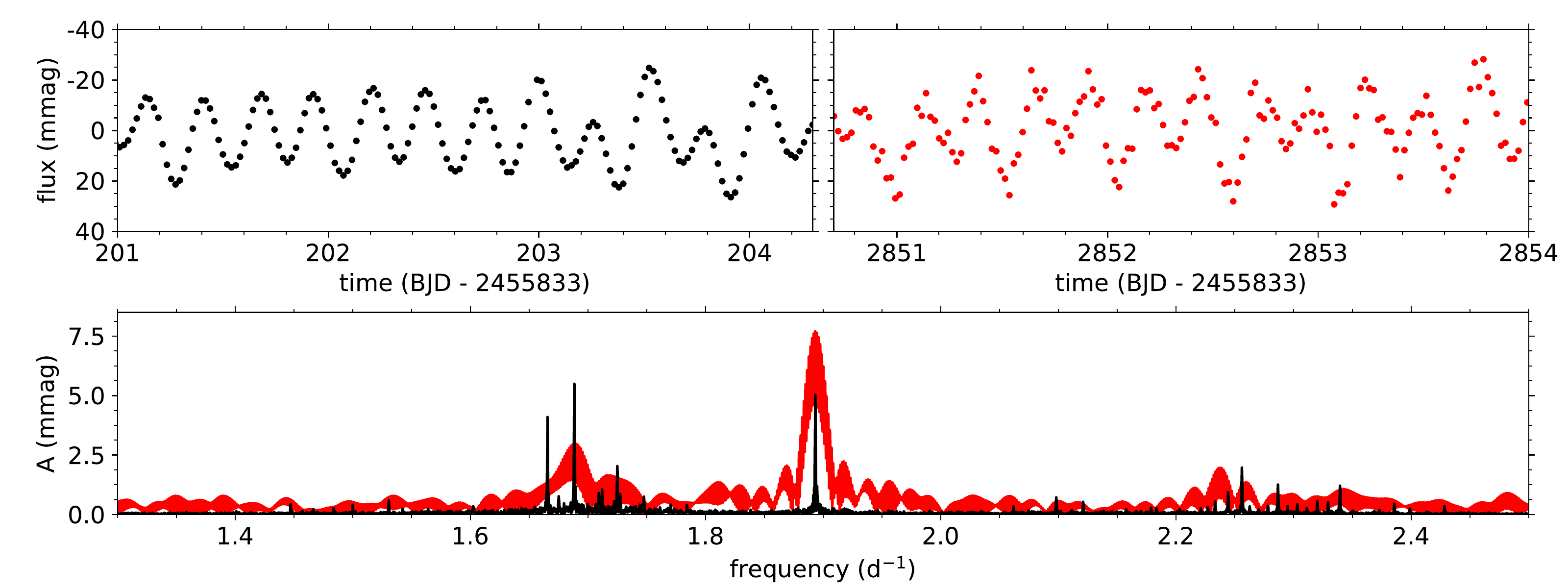}
    \caption{Comparison of the reduced {\em Kepler} (black) and TESS (red) light curves of KIC\,3341457, one the stars in our sample. {\em Top left:} Part of the {\em Kepler} light curve of KIC\,3341457. {\em Top right:} Part of the TESS light curve of KIC\,3341457. {\em Bottom:} Lomb-Scargle periodograms of the {\em Kepler} (black) and TESS (red) light curves of KIC\,3341457.\label{fig:kic3341457_tess-kepler}}
\end{figure*}

Despite these additional data reduction steps, the TESS data still do not reach the same quality level as the {\em Kepler} data, as illustrated in Fig.\,\ref{fig:kic3341457_tess-kepler}, and as a result have limited usage. Hence, throughout the rest of this work, the main analyses are based on the {\em Kepler} observations. The TESS observations are only used as a complementary source of information.

\section{Data analysis}
\label{sec:methods}
To identify targets with tidally perturbed g-mode pulsations, each of the light curves is analysed in the same systematic way. First, we model and subtract the binary signal from the light curve. Then we measure the frequencies of the independent g~modes from the residual light curve, and finally we identify which of these g~modes are tidally perturbed. 

\subsection{Binarity}
\label{subsec:binaries}
First, we model the observed variability associated with the binary motion. However, for most of the binaries in our sample, there are no published radial velocities or spectroscopic observations available. As a result, we do not have the required information to build detailed physical binary models \citep[e.g.][]{Southworth2020_bintable} for all stars in the sample. To ensure that all binaries are analysed homogeneously, we constructed a numerical model for the photometric binary variability using a sum of 100 sine waves \begin{equation}
    B\left(t\right) = \sum_{j=1}^{100} a_j\sin\left(2\pi\left(j\nu_{\rm orb} t + \varphi_j\right)\right),\label{eq:binary_model}
\end{equation}
where $\nu_{\rm orb}$ is the cyclic orbital frequency, and $a_j$ and $\varphi_j$ are the amplitude and phase of the $j^{\rm th}$ orbital harmonic, respectively. While this model is limited by the highest-order orbital harmonic that is considered, it does not smear out the binary signal \citep[see, e.g.][]{Bowman2019}, which could in turn modulate the observed pulsation signal. This risk does exist when the binary model is calculated by phase-folding and binning the light curve onto the orbit, as was done by for example \citet{Li2020_bin}. Because our aim was to study the relation between the binarity and g~modes, and we could benchmark our results against those presented by \citet{Li2020_bin}, the harmonic model in Eq.\,(\ref{eq:binary_model}) was best-suited to our work.

\subsection{Pulsation analysis}
\label{subsec:pulsations}
Following the harmonic modelling, we first used iterative prewhitening to determine the frequencies of the g-mode variability in the residual light curve, hereafter referred to as the pulsation light curve.  To minimise a potential bias of the measured g-mode pulsations caused by the partial occultation of the pulsation mode geometry, also called eclipse mapping \citep{Reed2001,Reed2005,Johnston2022} or spatial filtering \citep{Gamarova2003,Rodriguez2004,Gamarova2005}, we limit this analysis to the out-of-eclipse parts of the light curves. In each iterative step, we calculated the Lomb-Scargle periodogram \citep{Scargle1982} from the residual light curve, and fitted and subtracted a sine wave with the dominant frequency $\nu_j$ from the pulsation light curve. This was done for all frequencies with an optimal signal-to-noise ratio $S/N \geq 5.6$ based on statistical false-alarm simulations for {\em Kepler} data \citep{Baran2015,Zong2016,Bowman2021}, where $S/N$ was calculated as the ratio between the amplitude $a_j$ of the $j^{\rm th}$ fitted sine wave and the average amplitude of the residual Lomb-Scargle periodogram in a $1\,\rm d^{-1}$~window around the frequency of that sine wave. Next, we filtered the derived frequency list, based on both an automated selection, whereby only frequencies which were dominant in a $2.5\nu_{\rm res}$~window were accepted \citep[with $\nu_{\rm res}$ being the frequency resolution;][]{LoumosDeeming1978}, and subsequent manual inspection. From the resulting list of $N$ pulsation frequencies $\nu_j$, with corresponding amplitudes $a_j$ and phases $\varphi_j$, we built a light curve model
\begin{equation}
    L_c\left(t\right) = \sum_{j=1}^{N} a_j\sin\left(2\pi\left(\nu_j t + \varphi_j\right)\right).\label{eq:iterative_prew}
\end{equation}
This model $L_c(t)$ was then optimised with a non-linear fit to the pulsation light curve, using least-squares minimisation with the trust region reflective method, as implemented in the ``least\_squares'' routine in \texttt{lmfit}\footnote{\href{https://lmfit.github.io/lmfit-py/}{https://lmfit.github.io/lmfit-py/}} \citep{lmfit100}. To ensure that the solution had properly converged, the parameter values were visually inspected and compared with those from the input model. Finally, we redetermined the $S/N$~values associated with the different frequencies.

Next, we searched for g-mode patterns, using the methodology outlined by \citet{VanReeth2015_method}. We subsequently modelled these patterns with asymptotic g-mode series using the TAR, fitting the pulsation periods as a function of radial order \citep[as done in, e.g.][]{Li2020,VanReeth2022_HD112429} with the trust region reflective method, which allowed us to determine the mode geometries $(k,m)$ of the pulsations, the average near-core rotation rate $\nu_{\rm rot,nc}$ and the buoyancy travel time $\Pi_0$ of the pulsating star. Confidence intervals for $\nu_{\rm rot,nc}$ and $\Pi_0$ were calculated using a conservative $3-\sigma$ limit \citep{Takata2020a,VanReeth2022_HD112429}. These results were subsequently validated by comparing them with those from \citet{Li2020_bin}. 

\begin{figure*}
    \centering
    \includegraphics[width=\textwidth]{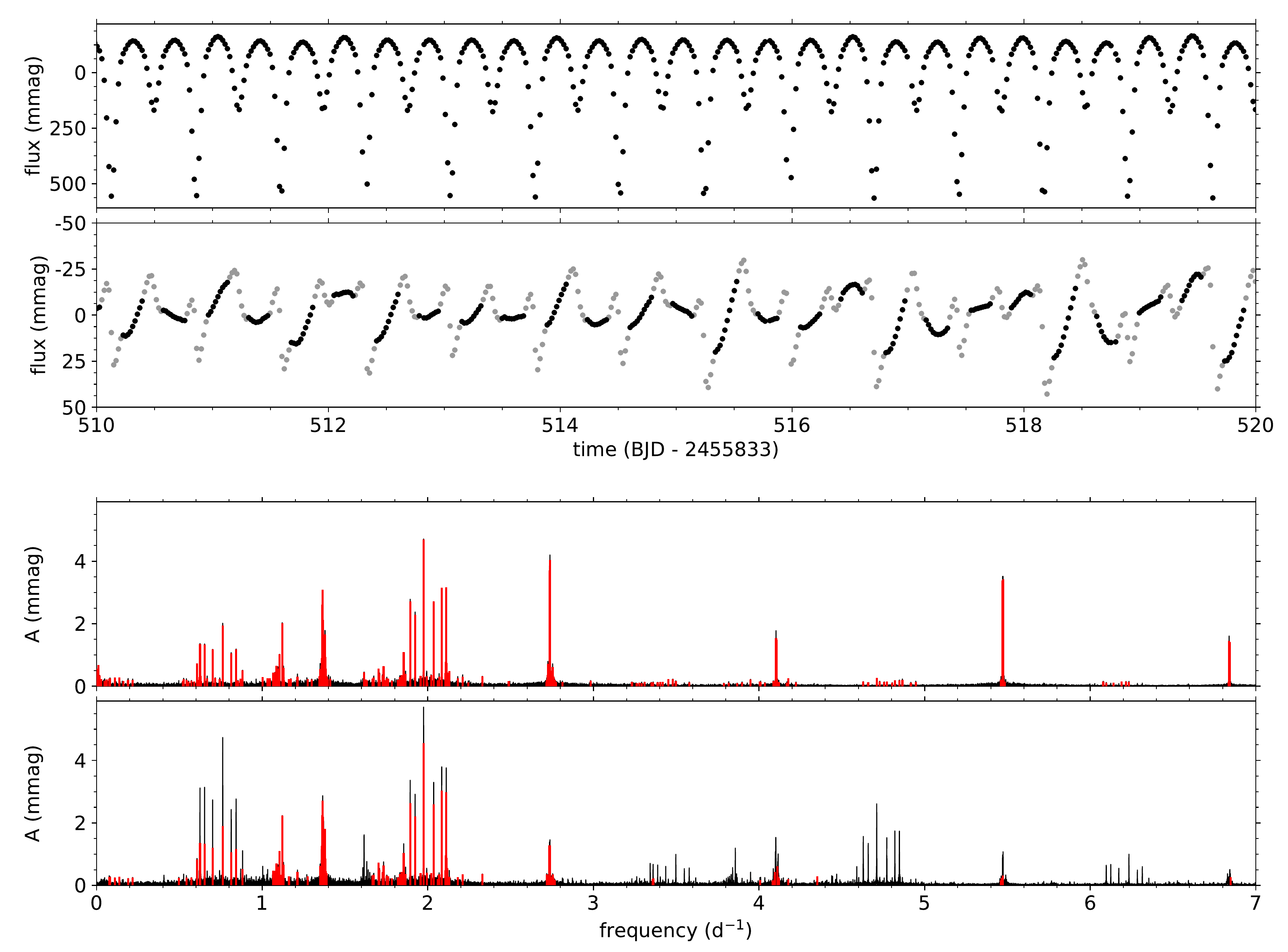}
    \caption{Different steps in the pulsation analysis of KIC\,3228863, one of the stars in our sample. {\em Top:} part of the reduced {\em Kepler} light curve. {\em $2^{\rm nd}$ panel:} part of the pulsation light curve, calculated by subtracting the harmonic model from the light curve, with the eclipses and the out-of-eclipse parts shown in grey and black, respectively. {\em $3^{\rm rd}$ panel:} Lomb-Scargle periodogram of the full pulsation light curve (black), with the prewhitened frequencies (red). {\em Bottom:} Lomb-Scargle periodogram of the out-of-eclipse parts of the pulsation light curve (black), with the prewhitened frequencies (red). \label{fig:kic3228863_eclipses}}
\end{figure*}

Different steps of this pulsation analysis are illustrated in Fig.\,\ref{fig:kic3228863_eclipses} for KIC\,3228863, one of the stars in our sample. As seen in the $2^{\rm nd}$ panel, there are strong residuals in the pulsation light curve during the eclipse phases, caused by eclipse mapping and light travel time variations \citep{Lee2014_V404Lyr,Lee2020_V404Lyr}. These can lead to the detection of additional frequencies during the iterative prewhitening routine, as shown in the $3{\rm rd}$ panel of Fig.\,\ref{fig:kic3228863_eclipses}, which are avoided by excluding the eclipse phases from the analysis, as illustrated in the bottom panel of Fig.\,\ref{fig:kic3228863_eclipses}.

\subsection{Tidal perturbations}
\label{subsec:tidal_pert}
In the next step, we identified which g-mode pulsations, if any, are tidally perturbed, by establishing if their frequencies $\nu_i$ are part of multiplets spaced by integer multiples of $\nu_{\rm orb}$. In this work, this criterion is fulfilled if there is at least one prewhitened frequency $\nu_j$ for which 
\begin{equation}
    |\nu_i - \nu_j - k\nu_{\rm orb}| \leq 3\sqrt{\sigma_{\nu,i}^2 + \sigma_{\nu,j}^2}~,
\end{equation} 
where $\sigma_{\nu,i}$ and $\sigma_{\nu,j}$ are the error margins on $\nu_i$ and $\nu_j$, and $k \in \mathbb{Z}$. For each of these candidate perturbed g~modes, we subsequently calculated the orbital-phase dependence of the modulations in two different ways. First, we reconstructed the tidal modulations from the detected $\nu_{\rm orb}$ spaced multiplets, using a semi-analytical model similar to the one developed by \citet{Jayaraman2022} as explained in Appendix\,\ref{sec:semi-analytical-perturbations}. Second, we phase-folded the pulsation light curve onto the binary orbit and divided the data points in 40 bins according to the orbital phase. Within each bin, we then refitted the light curve model $L_c(t)$ to the observations, optimising the pulsation amplitudes and phases but keeping the frequencies fixed. The solutions from both methods were visually inspected and required to be consistent with each other.

We note that our detection criteria for tidal perturbation are very strict. This allows us to present and discuss unambiguous detections of this phenomenon in Section\,\ref{sec:discussion}, but it is entirely possible that other stars in our sample also exhibit tidally perturbed pulsations, but did not fulfil our selection criteria.

\subsection{Rotational-modulation-like variability}
\label{subsec:rotmod}

Finally, many stars in our sample exhibit rotational modulation or rotational-modulation-like signals, often at or around the orbital harmonics of the binary signal. They meet the detection criteria for rotational modulation that were defined by \citet{VanReeth2018}, but they also have relatively high amplitudes and can thus be identified by eye as well. There are multiple possible physical origins, many of which are not related to rotation and not fully understood, as explained later in Sect.\,\ref{sec:obs-rotmod}. Physical modelling or a detailed understanding of this observed signal lies outside the scope of our work, but we provide an observational characterisation for the stars with detected tidally perturbed g~modes.

To characterise the rotational-modulation-like signal and its time dependence, we have evaluated its amplitude as a function of time and the binary orbital phase. To this end, we have selected 100-d sections of the pulsation light curve using a sliding box. For each section, we then removed the light curve model $L_c(t)$, which we presented in Eq.\,(\ref{eq:iterative_prew}), from the data, apart from the frequencies that are part of the rotational-modulation-like signal. For the stars in our sample, the rotational-modulation-like variability typically coincides with the orbital harmonics in the Lomb-Scargle periodogram, and measured frequencies are considered part of the signal if they lie within a 0.05\,$\rm d^{-1}$-window, centred on one of the orbital harmonics. This window width is sufficiently broad to cover the unresolved rotational-modulation-like peaks in the periodogram, but not overlap with the g-mode frequency ranges. We then phase-folded the residual light curve section with the orbital period, divided the data into 40 bins and calculated the median flux within each bin. To validate our results, we repeated this analysis for 200-d sections of the pulsation light curve, using measured pulsation frequencies with $S/N \geq 4.0$ and 0.1\,$\rm d^{-1}$-windows around the rotational-modulation-like peaks in the periodogram. We then compared these results to ensure that they were not affected by any residual pulsation or harmonic signal, but for clarity the detailed discussion in Sect.\,\ref{sec:obs-rotmod} is limited to our first set of results.

\section{Discussion of individual systems}
\label{sec:discussion}
Following the data analysis methodology described in the previous section, we detected tidally perturbed g-mode pulsations in 5 of the 35 sample targets of \citet{Li2020_bin}. An overview of their derived parameter values is listed in Table \ref{tab:tidal-perturb-overview}, and each of them is discussed in detail below. Additional figures that illustrate the detections and complement the discussions below, are included in Appendix\,\ref{appendix:tidal-detections}.

\begin{table*}
	\centering
	\caption{Overview of the measured parameter values for our binary stars with detected $\gamma$\,Dor-type pulsations and tidal perturbations.}
	\label{tab:tidal-perturb-overview}
	\begin{tabular}{lllllll} 
		\hline\hline\vspace{-9pt}\\
                                   & KIC\,3228863    & KIC\,3341457 & KIC\,4947528  & KIC\,9108579    & KIC\,12785282   & V456\,Cyg${}^{(1)}$\\
		\hline\vspace{-9pt}\\
    TIC                            & 394179296       & 137094904    & 169560376     & 273583841       & 406997952       & 15795549           \\
    eclipsing                      & y               & n            & n             & n               & y               & y                  \\
    $\mathrm{m}_{Kepler}$ (mag)    & 11.816          & 13.873       & 13.910        & 11.560          & 13.518          & -             \\
    $\nu_{\rm orb}$ ($\rm d^{-1}$) & 1.36809332(4)   & 1.8932976(5) &  2.0121190(1) &  0.85497202(16) &  1.26782902(13) &  1.122091(8)       \\
    $\nu_{\rm rot,nc}$ ($\rm d^{-1}$) &     1.371(3)    &     1.865(1) &      2.03(2)  &        0.847(7) &        1.264(3) &     1.113(5)       \\
    $\Pi_0$ (s)                    &    4370(100)    &     3870(40) &    4390(310)  &      4240(160)  &        4214(30) &     4300(50)       \\
    pulsating component            &      primary    &      unknown &       unknown &         unknown &         primary &    secondary       \\
        \hline\vspace{-9pt}\\
    $k$                            &             0   &           0  &            0  &              0  &               0 &           0        \\
    $m$                            &             1   &           1  &            1  &              1  &               1 &           1        \\
    $n_{\rm min}$                  &           -61   &         -72  &          -42  &            -47  &             -56 &         -31        \\
    $n_{\rm max}$                  &           -28   &         -41  &          -21  &            -40  &             -19 &         -18        \\
    $n_{\rm span}$                 &            34   &          32  &           22  &              8  &              38 &          14        \\
    tidal perturbations            &             y   &           y  &            y  &              n  &               y &           y        \\
        \hline\vspace{-9pt}\\
    $k$                            &            -2   &          -2  &       -  &              0  &          - &      -        \\
    $m$                            &            -1   &          -1  &       -  &              2  &          - &      -        \\
    $n_{\rm min}$                  &           -24   &         -53  &       -  &            -61  &          - &      -        \\
    $n_{\rm max}$                  &           -18   &         -33  &       -  &            -49  &          - &      -        \\
    $n_{\rm span}$                 &             7   &          21  &       -  &             13  &          - &      -        \\
    tidal perturbations            &             n   &           y  &       -  &              y  &          - &      -        \\
 		\hline
	\end{tabular}
	\tablefoot{The middle and bottom sections contain the relevant parameters for the first and (if detected) second g-mode period-spacing pattern of each target, respectively. The $\nu_{\rm rot,nc}$ parameter is the near-core rotation frequency, measured from asteroseismic modelling. The $n_{\rm span}$ parameter indicates the total number of radial orders that the g-mode pulsations within the detected period-spacing patterns span. The ``tidal perturbations'' parameter indicates when tidal perturbations were detected for any pulsation in a g-mode period-spacing patterns.}
\tablebib{(1)~studied by \citet{VanReeth2022a}}
\end{table*}

\subsection{KIC\,3228863}
\label{subsec:kic03228863}

\begin{figure}
    \centering
    \includegraphics[width=\columnwidth]{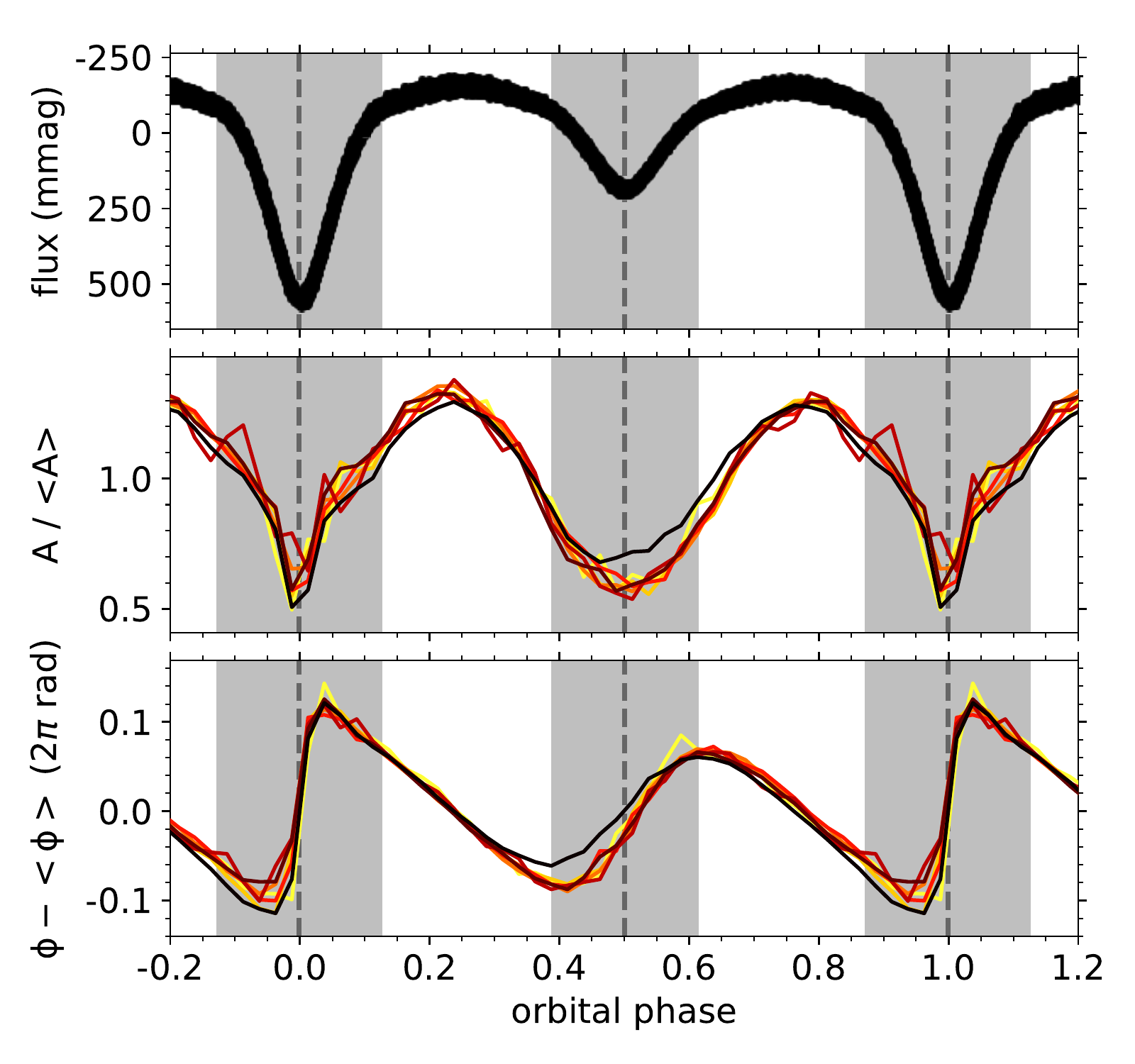}
         \caption{Relative tidal perturbation of all selected $(k,m) = (0,1)$ g~modes of KIC\,3228863. {\em Top:} the phase-folded {\em Kepler} light curve of KIC\,3228863 (black). The eclipses are marked in grey, with the dashed lines indicating the centres of the eclipses. {\em Middle:} pulsation amplitude modulations as a function of the binary orbital phase, calculated numerically within separate orbital phase bins and divided by the average intrinsic mode amplitudes. Different line colours scale with the average intrinsic mode amplitudes, from high (black) to low (yellow). The binary eclipses are marked in grey. {\em Bottom:} differential pulsation phase modulations as a function of the binary orbital phase, calculated numerically from binned data, with $\langle\phi\rangle$ indicating the average intrinsic mode phase.}
    \label{fig:kic03228863_allpuls}
\end{figure}

KIC\,3228863 is one of the two eclipsing binary systems in our sample, as can be seen from the phase-folded light curve in the top panel of Fig.\,\ref{fig:kic03228863_allpuls}, where the grey areas indicate the orbital phases of the eclipses. It is a semi-detached binary, part of a quadruple system, and our only target with published radial velocities and orbital parameters \citep{Lee2014_V404Lyr,Lee2020_V404Lyr}.
After the frequency analysis of the pulsation light curve from the {\em Kepler} space mission, we detected and modelled two period-spacing patterns for g-mode pulsations with mode identification $(k,m) = (0,1)$ and $(-2,-1)$, respectively, in agreement with the results obtained by \citet{Li2020_bin}. It shows that the near-core region of the primary is synchronously rotating ($\nu_{\rm rot,nc} = 1.371(3)\,\rm d^{-1}$) with the binary orbit ($\nu_{\rm orb} = 1.36809332(4)\,\rm d^{-1}$) within the confidence intervals of the measured frequencies.

Using our strict criteria, explained in the previous section, we detected tidal perturbation of the seven most dominant $(k,m) = (0,1)$~modes. There are indications that at least some of the remaining independent g-mode pulsations, including those with $(k,m) = (-2,-1)$, are also tidally perturbed, but these were not formally detected. Because our aim is to provide a clean detection and characterisation of tidally perturbed g~modes, these remaining oscillations are therefore not discussed here. 

The observational signatures of the detected tidal perturbed pulsations are illustrated in Figs.\,\ref{fig:kic03228863_allpuls} and \ref{fig:kic03228863_period-spacings} to \ref{fig:kic03228863_bestpuls}. In these figures, we see a decrease (increase) in the observed pulsation amplitude during the primary (secondary) eclipse, as well as a saw-tooth signature in the pulsation phase modulation during the primary eclipse. These signatures are caused by eclipse mapping \citep{Reed2001,Reed2005,Johnston2022}, the partial occultation of the pulsation mode geometry, and indicate that the g-mode pulsations belong to the primary. This is in agreement with the conclusions from \citet{Lee2020_V404Lyr}, which were based on the locations of the binary components in the HR~diagram. Interestingly, the correlation between the amplitude and phase modulations is qualitatively similar both during and out of the eclipses: when the pulsation amplitude is maximal, the pulsation phase decreases as a function of the orbital phase, and vice versa. Moreover, the observed tidal modulations of the seven dominant pulsations are similar. This is illustrated in Fig.\,\ref{fig:kic03228863_allpuls}, where we compare their relative amplitude and phase modulations, as calculated from the binned data analysis. Here, the modulations of the dominant pulsation are indicated in black, while the modulations for the smallest pulsation are shown in yellow. With the exception of the dominant pulsation, the relative modulations of the different pulsations agree with each other within their uncertainties.

\subsection{KIC\,3341457}
\label{subsec:kic03341457}
\begin{figure}
    \centering
    \includegraphics[width=\columnwidth]{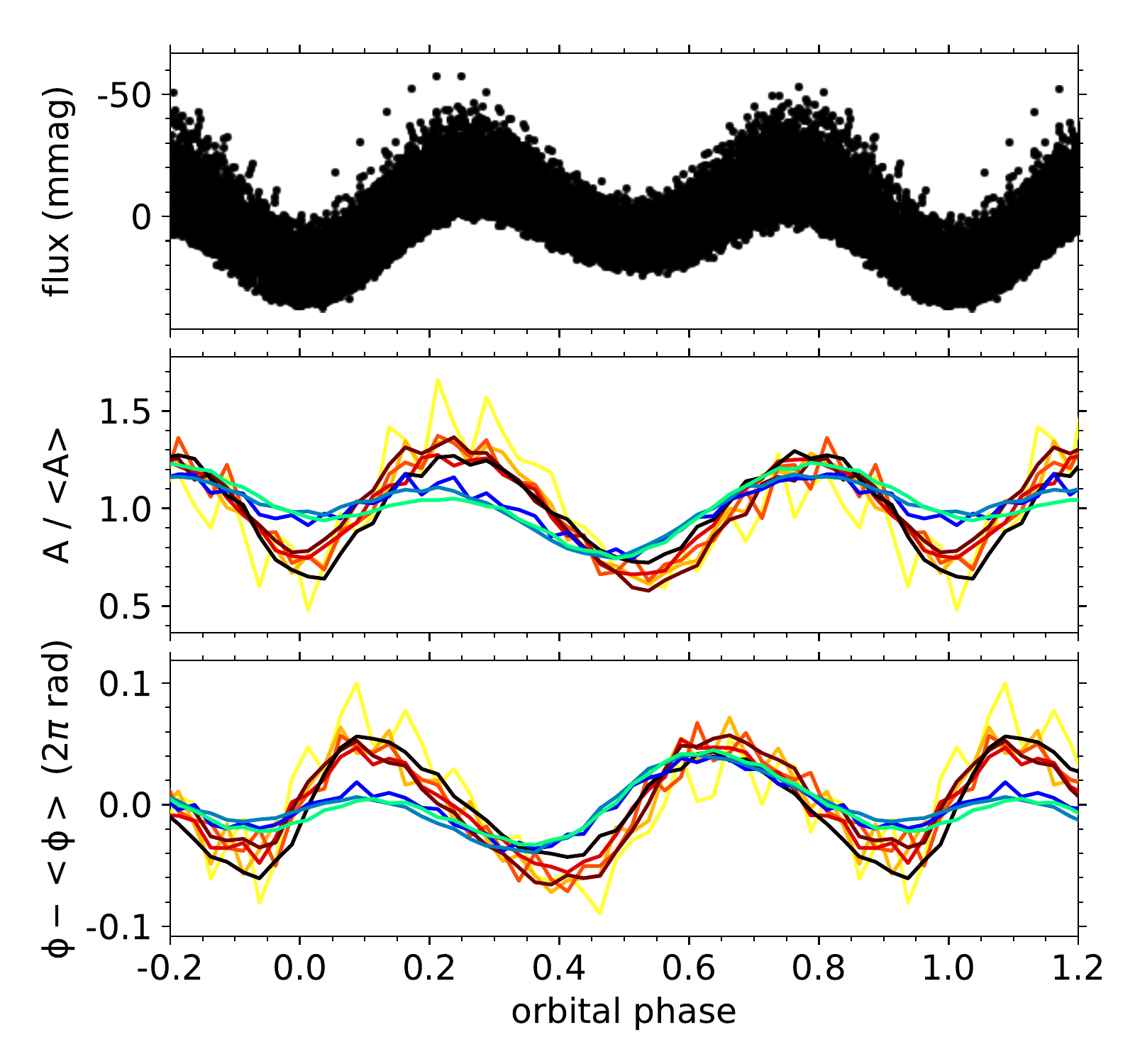}
    \caption{Relative tidal perturbation of all selected g and r~modes of KIC\,3341457. {\em Top:} the phase-folded {\em Kepler} light curve of KIC\,3341457 (black). {\em Middle:} relative pulsation amplitude modulations as a function of the binary orbital phase, similar to the middle panel of Fig.\,\ref{fig:kic03228863_allpuls}. Modulations shown using the black-red-yellow colour scale correspond to $(k,m) = (0,1)$ g~modes. The blue-green colour scale is used for modulations of $(k,m) = (-2,-1)$ r~modes. {\em Bottom:} differential pulsation phase modulations as a function of the binary orbital phase, similar to the bottom panel of Fig.\,\ref{fig:kic03228863_allpuls}. }
    \label{fig:kic03341457_allpuls}
    \includegraphics[width=\columnwidth]{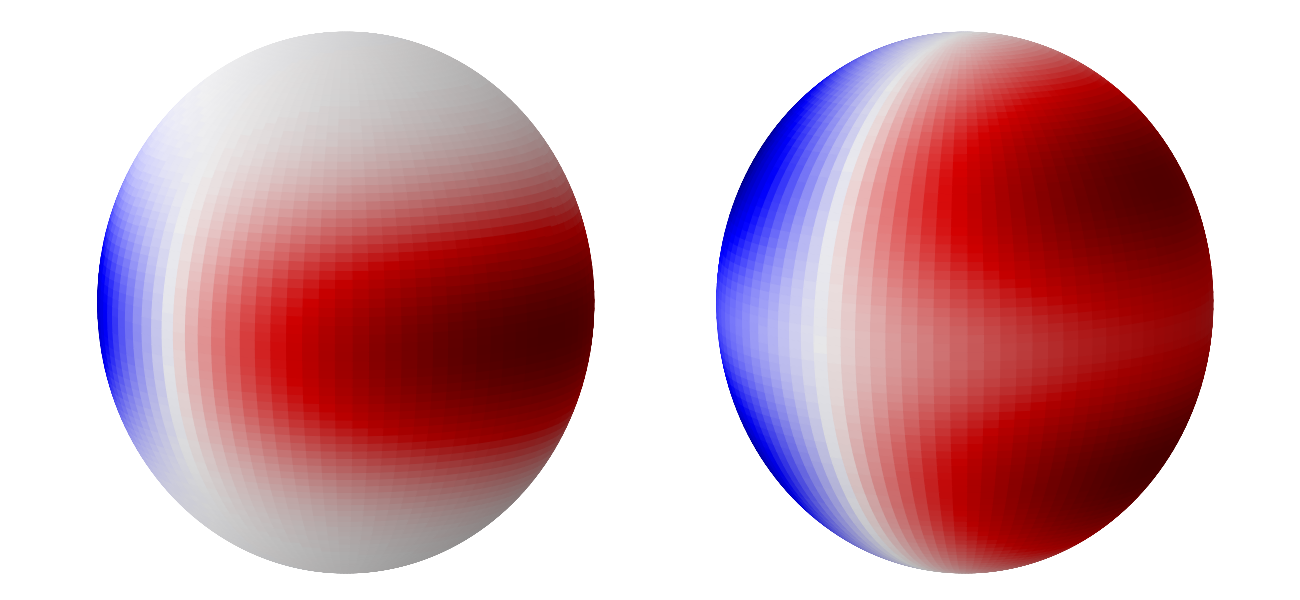}
    \caption{Geometries of the dominant $(k,m) = (0,1)$ g~mode (left) and $(k,m) = (-2,-1)$ r~mode of KIC\,3341457 (right), ignoring the effects of tidal deformation and perturbation. Regions on the stellar surface where the pulsations cause the temperature to decrease (increase), are indicated in red (blue). \label{fig:kic03341457_modegeometries}}
\end{figure}

Contrary to KIC\,3228863, we cannot assign the pulsations of KIC\,3341457 to a specific component. This is a non-eclipsing, low-amplitude ellipsoidal variable, as can be seen from the top panel of Fig.\,\ref{fig:kic03341457_allpuls}. Because there are no eclipses, it is not possible to determine from the light curve which binary component is pulsating. Moreover, as there are no time series of high-resolution spectroscopy available for this target yet, we cannot place the binary components on the Hertzsprung-Russell diagram or measure their masses.

However, in agreement with \citet{Li2020_bin}, we detected two g-mode patterns among the prewhitened pulsation frequencies, with mode identification $(k,m) = (0,1)$ and $(-2,-1)$, respectively. From the asteroseismic analysis of these patterns, we found that the pulsating component is clearly rotating a bit slower (near-core rotation $\nu_{\rm rot,nc} = 1.865(1)\,\rm d^{-1}$) than the binary orbit ($\nu_{\rm orb} = 1.8932976(5)\,\rm d^{-1}$), making this the only target in our sample with clear asynchronous rotation. 

Moreover, as shown in the middle and bottom panels of Fig.\,\ref{fig:kic03341457_allpuls}, we have detected tidal perturbations of three r~modes (with $(k,m) = (-2,-1)$) as well as six prograde dipole modes ($(k,m) = (0,1)$). While the modulation morphology is again very similar for g~modes with identical mode identification $(k,m)$, there are clear differences for pulsations with different mode geometries. While pulsations with $(k,m) = (0,1)$ are (mostly) spheroidal and dominant near the equator, the r~modes with $(k,m) = (-2,-1)$ are toroidal oscillations with dominant temperature variations at latitudes well above and below the equator \citep{Saio2018}, as illustrated in Fig.\,\ref{fig:kic03341457_modegeometries}. As a result, both types of pulsations show different levels of sensitivity to the tidal deformation of the star, and as shown in Figs.\,\ref{fig:kic03341457_bestgpuls} and \ref{fig:kic03341457_bestrpuls}, they have different morphologies as a function of the binary orbital phase. This proves that the pulsations do not merely scale with the tidal deformation of the pulsating star, but are effectively distorted.

\subsection{KIC\,4947528}
\label{subsec:kic04947528}
\begin{figure}
    \centering
    \includegraphics[width=\columnwidth]{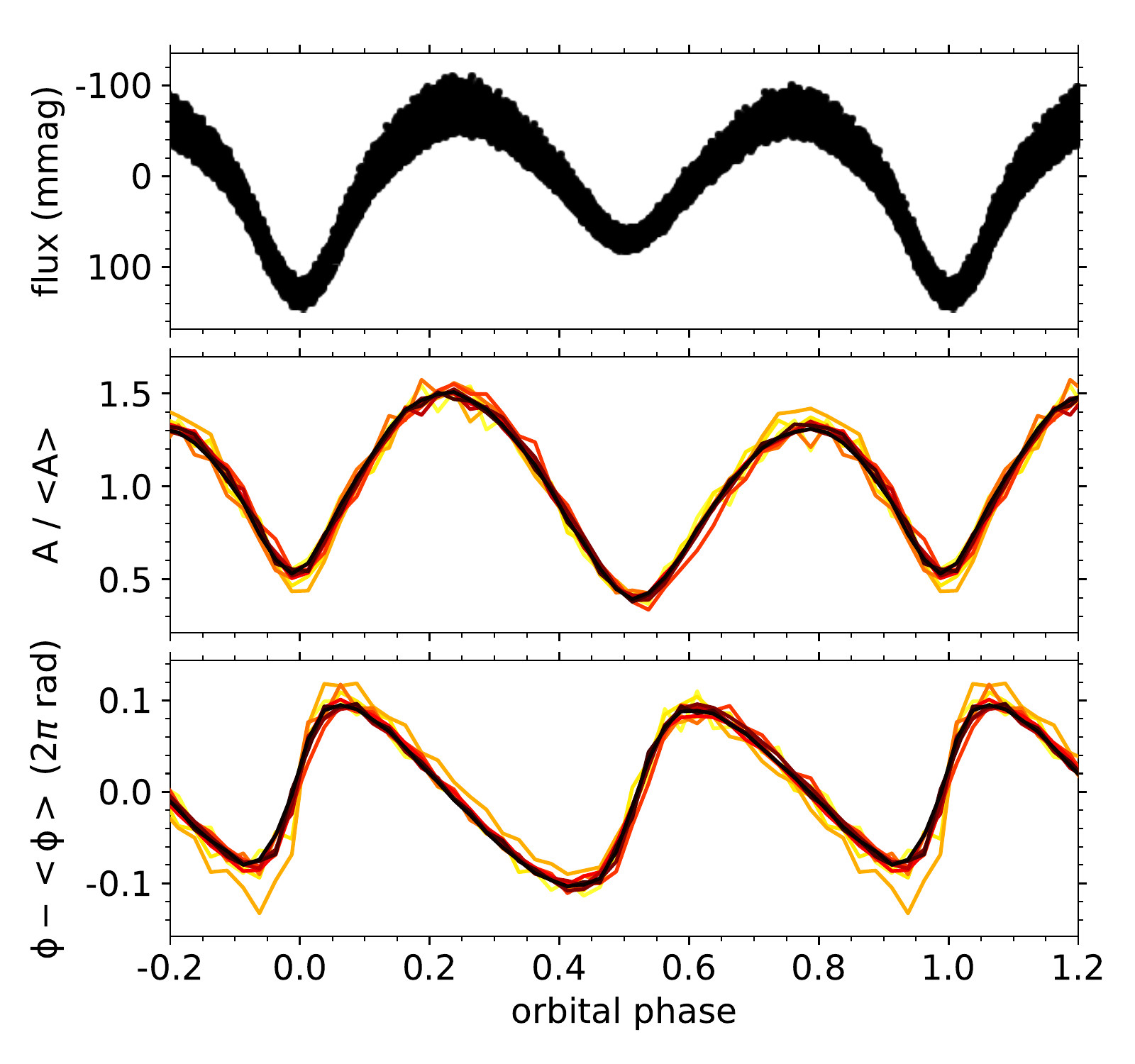}
    \caption{Relative tidal perturbation of all selected $(k,m) = (0,1)$  g~modes of KIC\,4947528, similar to those shown in Fig.\,\ref{fig:kic03228863_allpuls}.\label{fig:kic04947528_allpuls}}
\end{figure}

KIC\,4947528 is the shortest-period binary in our sample. Similar to KIC\,3341457, we do not have spectroscopic data and cannot identify which component is pulsating. However, based on our asteroseismic results we can confirm that the pulsating star is rotating synchronously (near-core rotation $\nu_{\rm rot,nc} = 2.03(2)\,\rm d^{-1}$) with the binary orbit ($\nu_{\rm orb} = 2.0121190(1)\,\rm d^{-1}$) within the confidence intervals of the measured frequencies. The ten most dominant g~modes, with $(k,m) = (0,1)$, are clearly tidally perturbed, as illustrated in Figs.\,\ref{fig:kic04947528_allpuls} and \ref{fig:kic04947528_period-spacings} to \ref{fig:kic04947528_bestpuls}. The relative pulsation amplitude and phase modulations again agree with each other within their uncertainties, and similar to our observations for KIC\,3228863 and KIC\,3341457, the amplitudes and phases are correlated with each other.

\subsection{KIC\,9108579}
\label{subsec:kic09108579}
KIC\,9108579 is the only binary system in our selection with an orbital period of more than a day ($\nu_{\rm orb} = 0.85497202(16)\,\rm d^{-1}$). Just as for KIC\,3341457 and KIC\,4947528, the pulsating component in the system could not be identified. However, from our analysis of the detected pulsations, we again found that the pulsating star is rotating synchronously (near-core rotation $\nu_{\rm rot,nc} = 0.847(7)\,\rm d^{-1}$)  within the confidence interval of the measured rotation frequency.

\begin{figure}
    \centering
    \includegraphics[width=\columnwidth]{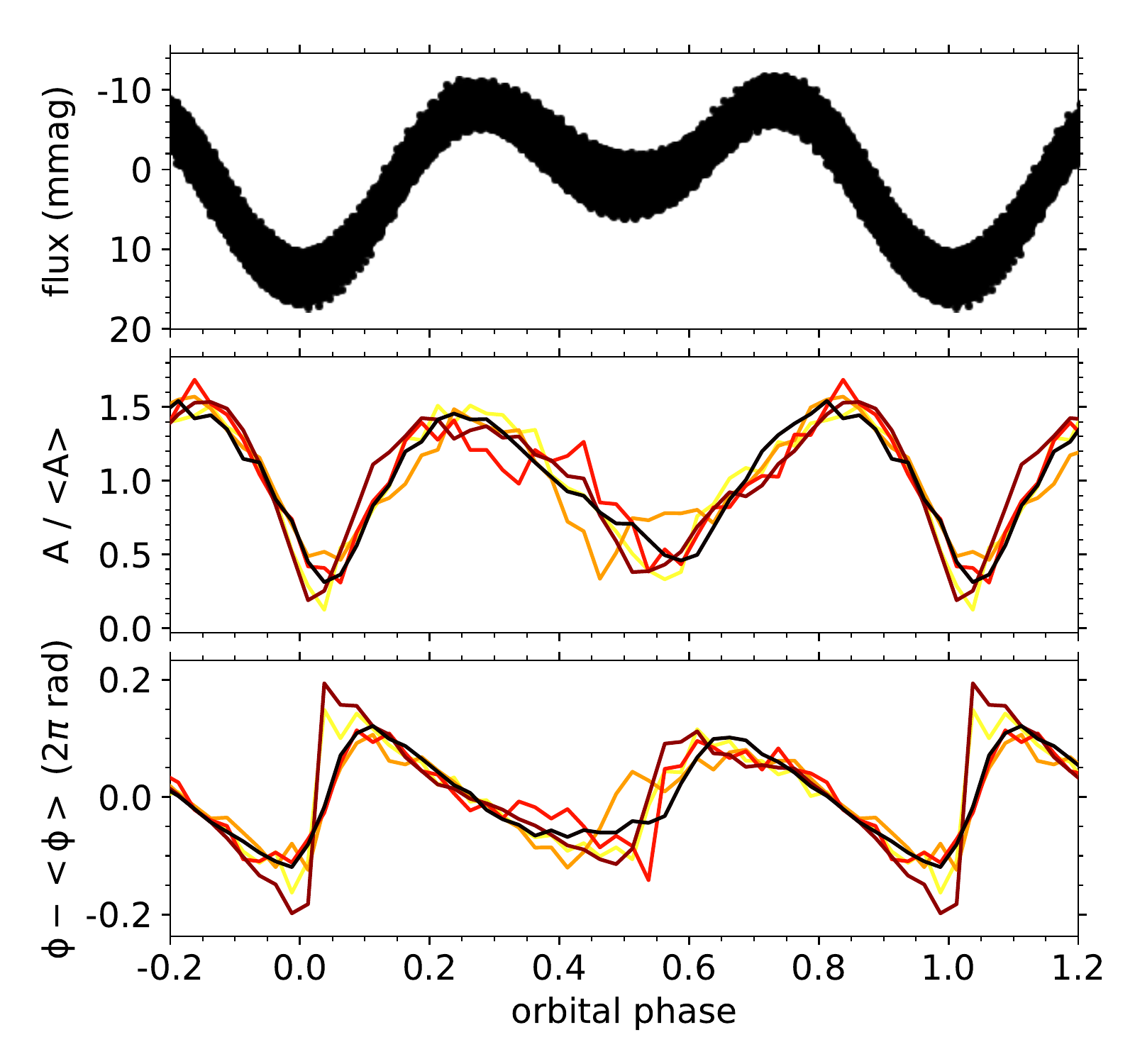}
    \caption{Relative tidal perturbation of all selected $(k,m) = (0,2)$  g~modes of KIC\,9108579, similar to those shown in Fig.\,\ref{fig:kic03228863_allpuls}.\label{fig:kic09108579_allpuls}}
\end{figure}

KIC\,9108579 is the only target in our selection to clearly exhibit both prograde dipole ($(k,m) = (0,1)$) and quadrupole ($(k,m) = (0,2)$) g~modes. Interestingly, while we could not formally detect tidal perturbations for any of the dipole modes, the five most dominant quadrupole modes exhibit tidal perturbations that are qualitatively similar to the tidal perturbations observed for the previous selected stars, as illustrated in Figs.\,\ref{fig:kic09108579_allpuls} and \ref{fig:kic09108579_period-spacings} to \ref{fig:kic09108579_bestpuls}. Within their confidence intervals, the relative amplitude and phase modulations of the different modes are again almost identical. We also note that, although the morphology of the amplitude modulations is similar to the ellipsoidal flux variations of the binary, which is comparable to the other binaries and as shown in Fig.\,\ref{fig:kic09108579_allpuls}, there is a large offset in orbital phase between them. While the ellipsoidal variability reaches its minima at orbital phases of 0.0 and 0.5, the minima of the amplitude modulations of the dominant pulsation are at orbital phases of 0.057 and 0.557.

\subsection{KIC\,12785282}
\label{subsec:kic12785282}
\begin{figure}
    \centering
    \includegraphics[width=\columnwidth]{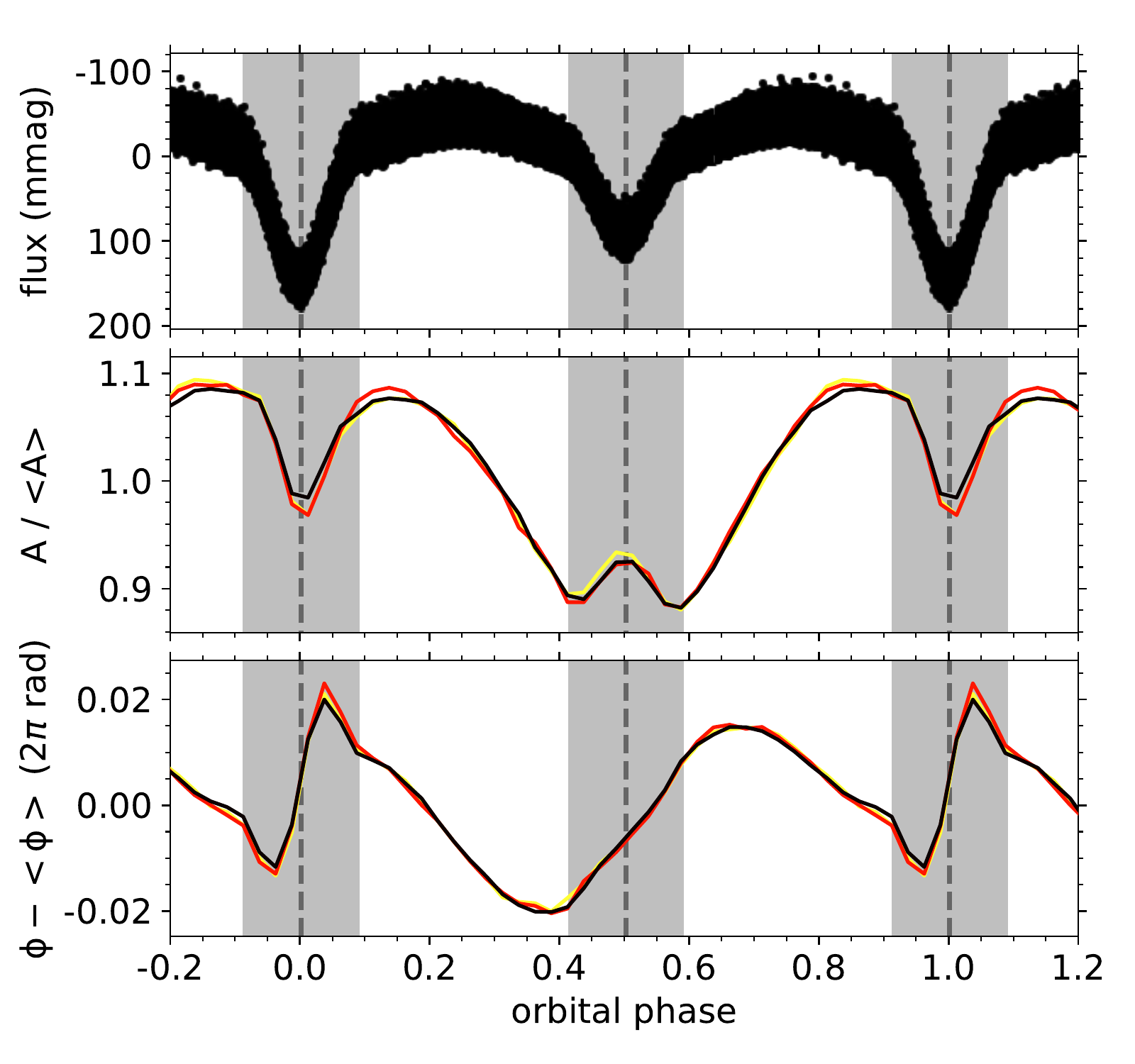}
    \caption{Relative tidal perturbation of all selected $(k,m) = (0,1)$  g~modes of KIC\,12785282, similar to those shown in Fig.\,\ref{fig:kic03228863_allpuls}.\label{fig:kic12785282_allpuls}}
\end{figure}

Our final target, KIC\,12785282, is the second eclipsing binary system among our selected targets. Contrary to \citet{Li2020_bin}, who detected a g-mode period-spacing pattern with mode identification $(k,m) = (0,1)$ spanning five radial orders, we found a much longer pattern spanning 38 radial orders. This allowed us to place much stronger constraints on the near-core rotation rate ($\nu_{\rm rot,nc} = 1.264(3)\,\rm d^{-1}$) of the pulsating star, and to confirm that it is also rotating synchronously ($\nu_{\rm orb} = 1.26782902(13)\,\rm d^{-1}$) within the confidence intervals of the measured frequencies. However, in spite of the large number of identified pulsation modes in our period-spacing pattern, only the three dominant modes fulfilled our formal criteria for the detection of tidal perturbation. As illustrated in Figs.\,\ref{fig:kic12785282_allpuls} and \ref{fig:kic12785282_period-spacings} to \ref{fig:kic12785282_bestpuls}, their behaviour is consistent with what we have observed for our other selected stars. Moreover, the pulsation amplitude and phase modulations during the eclipses demonstrate that the primary component is the pulsating star, and that the g~modes are amplified in the hemisphere facing the companion. This is similar to what was observed for V456\,Cyg \citep{VanReeth2022a}, although in that case the pulsating star was the secondary.

\section{Characterisation of the tidally perturbed g-modes}
\label{sec:characterisation}
Following the detailed individual analysis of all targets with tidally perturbed g-mode pulsations, there are clear similarities between these binary systems. They are very close, quasi-synchronised, circularised \citep{Li2020_bin,VanReeth2022a}, and have an orbital period of the order of or less than a day. Moreover, the tidal g-mode perturbations themselves also have clear common features. Unfortunately, the number of tidally perturbed stars has proven to be too small for a successful statistical sample analysis, but even a qualitative description provides very useful insights into the physics behind them. First, the relative tidal perturbations of pulsations with the same mode identification are often almost identical. In other words, the perturbations are nearly independent of the intrinsic pulsation mode amplitudes, radial orders, and phases. However, tidal perturbations of g-mode pulsations with different mode identification $(k,m)$ within the same star can be very dissimilar, and there are clear observational differences between the stars as well.

\begin{figure}
    \centering
    \includegraphics[width=\columnwidth]{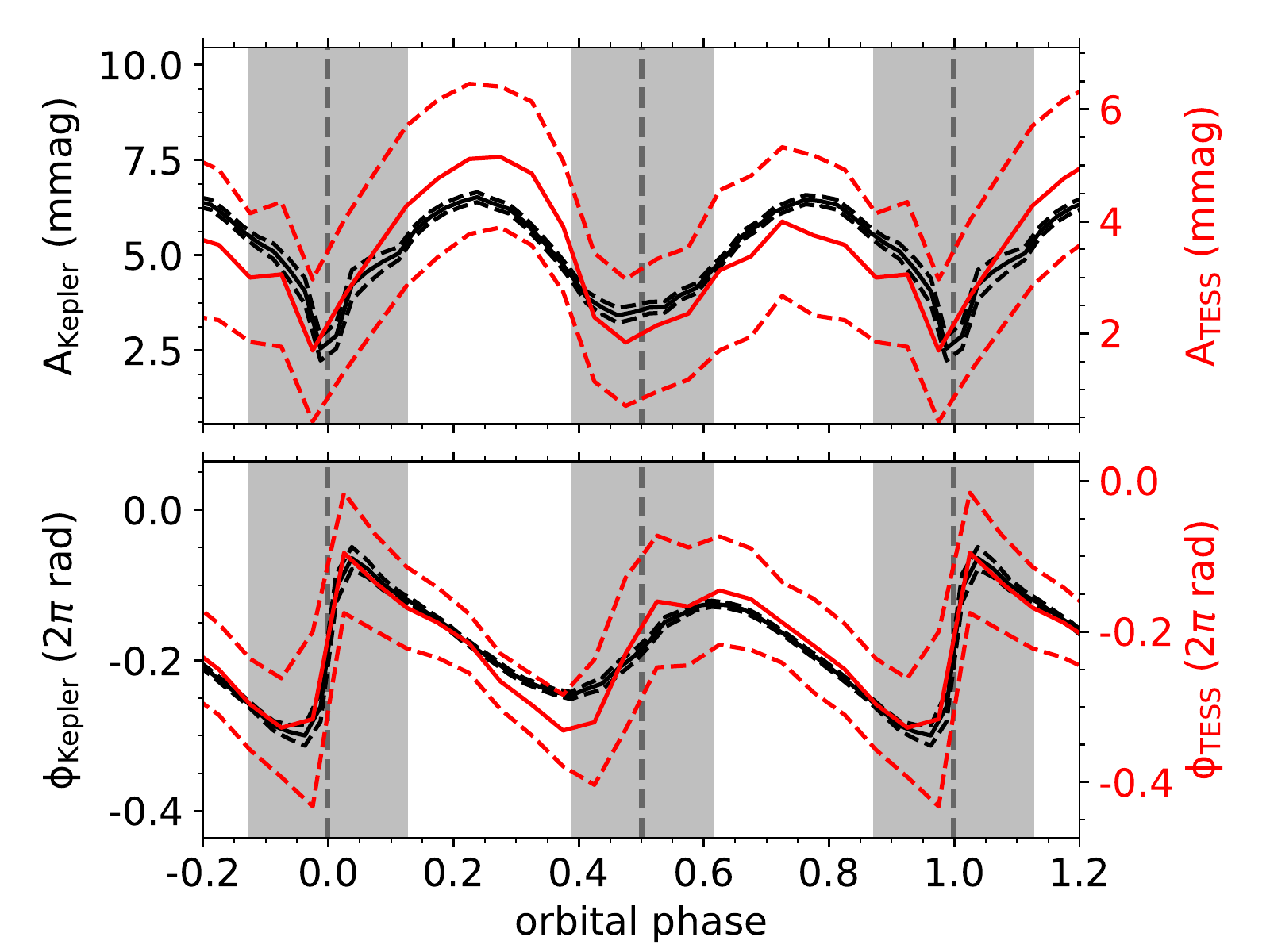}
    \caption{Comparison of the observed modulations of the dominant $(k,m) = (0,1)$ g~mode of KIC\,3228863 (with $\nu = 1.974605(3)\,\rm d^{-1}$) in the available {\em Kepler} (black) and TESS photometry (red).\label{fig:kic03228863_kepler-tess}}
\end{figure}

\begin{figure}
    \centering
    \includegraphics[width=\columnwidth]{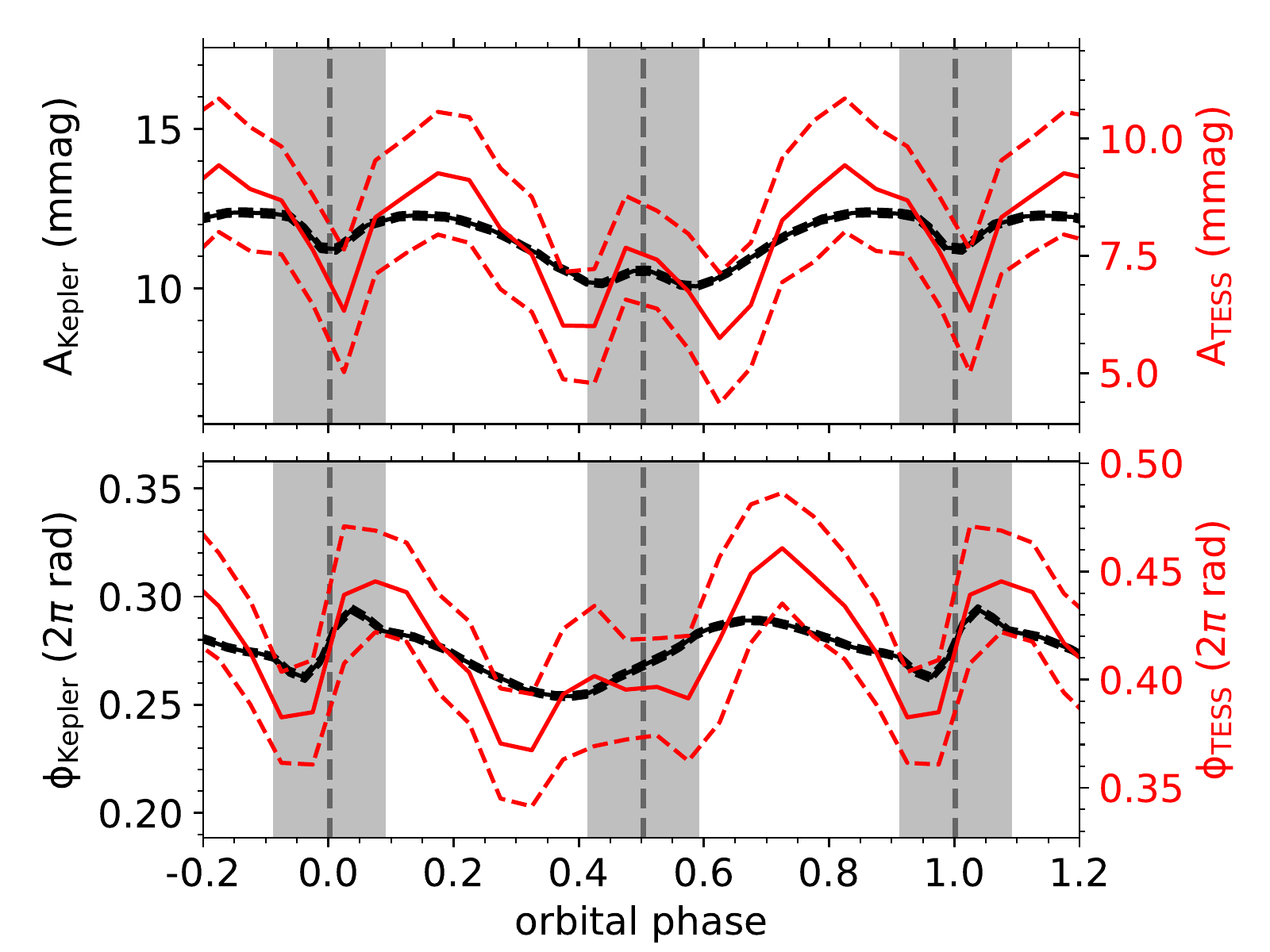}
    \caption{Comparison of the observed modulations of the dominant $(k,m) = (0,1)$ g~mode of KIC\,12785282 (with $\nu = 1.8971606(7)\,\rm d^{-1}$) in the available {\em Kepler} (black) and TESS photometry (red).\label{fig:kic12785282_kepler-tess}}
\end{figure}

Second, the observed tidal perturbations reach maximal amplitudes once or twice per orbital cycle, and are often strongly correlated with the light variations caused by the ellipsoidal variability of the system. However, these correlations are not exact, indicating that the tidal g-mode perturbations do not merely scale with the deformation of the pulsating star. Third, for all perturbed g~modes, the phase modulations decrease (increase) as a function of the binary orbital phase when the amplitude modulations are maximal (minimal), and the pulsation phase variations are relatively small and less than 0.5\,rad. Fourth, within the limits of the available data, the relative tidal g-mode perturbations are found to be independent of the chosen photometric passband. This is illustrated in Figs.\,\ref{fig:kic03228863_kepler-tess} and \ref{fig:kic12785282_kepler-tess}, where we compare the TESS observations of the dominant tidally perturbed modes of KIC\,3228863 and KIC\,12785282 with the {\em Kepler} data. However, because of the limited amount of available TESS data, the tidal perturbations were not detectable for any of the other pulsations or stars, and the confidence intervals on these two observed pulsations are large. 

\subsection{Pulsation amplitude and phase relations} 
\label{subsec:ampl-phase}
\begin{figure}
    \centering
    \includegraphics[width=\columnwidth]{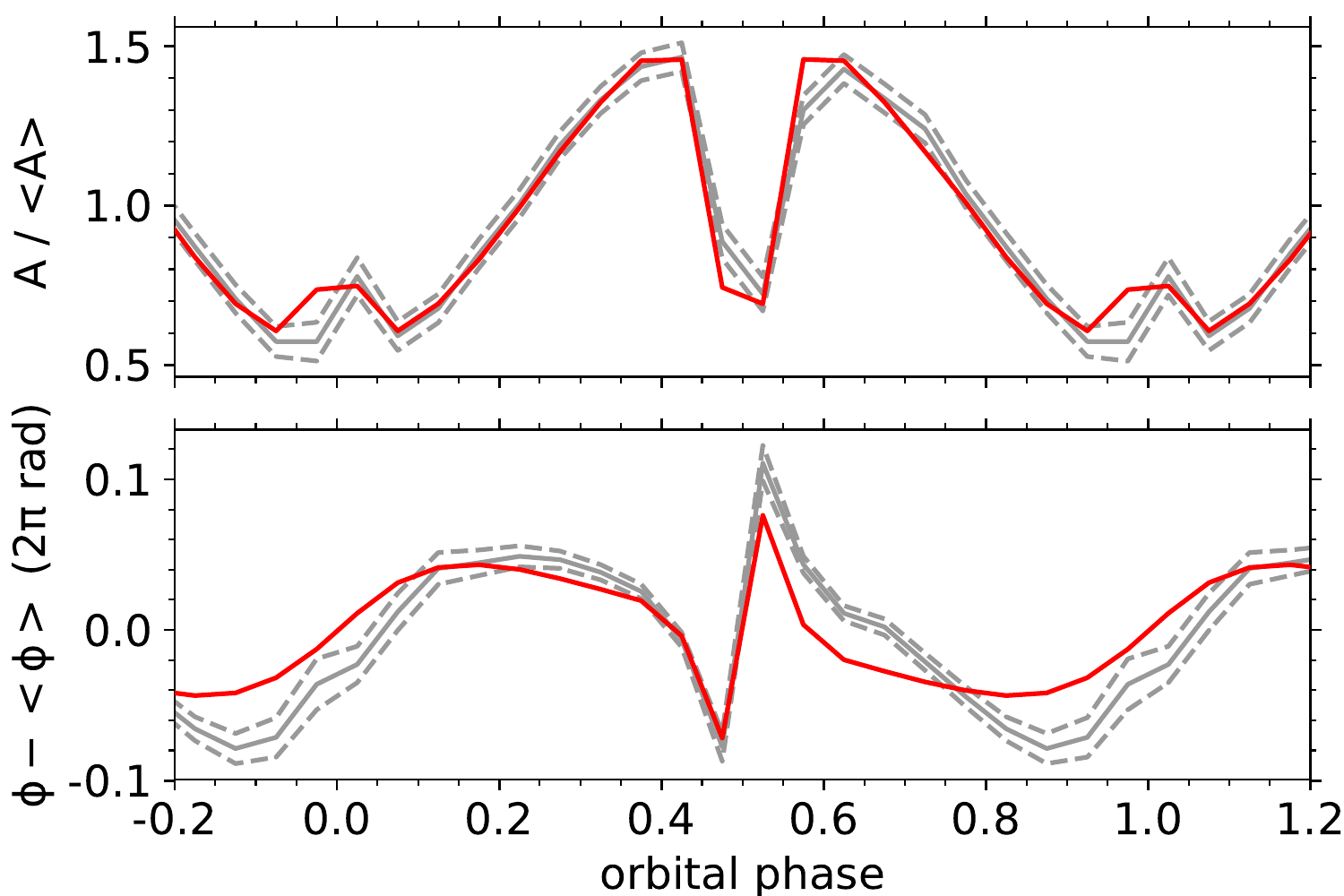}
    \caption{\label{fig:obs-V456Cyg_model}Comparison of the dominant tidally perturbed pulsation of V456\,Cyg \citep[light grey;][]{VanReeth2022a} with our toy model (red; described in Appendix \ref{appendix:toymodel}). {\em Top:} the observed amplitude modulations as a function of the orbital phase. {\em Bottom:} the observed pulsation phase modulations as a function of the orbital phase, caused by the geometry of the pulsation amplitude distortion across the stellar surface.}
\end{figure}

Despite the different characteristics of our stars, there are obvious relations between the amplitude and phase modulations of the tidally perturbed pulsations. However, as the measured pulsation phases depend on multiple aspects like intrinsic pulsation phase variations, the light travel time from the pulsating star to the telescope \citep[e.g.][]{Murphy2014}, and geometric effects such as spatial filtering \citep{Gamarova2003,Rodriguez2004,Gamarova2005} or tidal tilting \citep[e.g.][]{Fuller2020}, investigating this relation can be difficult. Hence, we have built a toy model of a g~mode with a distorted pulsation amplitude in a synchronised, circular binary system, as described in detail in Appendix \ref{appendix:toymodel}, deliberately ignoring all other effects. In short, we used the parameter values from V456\,Cyg \citep[obtained by][]{VanReeth2022a} to construct the binary model, included the dominant $(k,m) = (0,1)$ g-mode pulsation of the secondary component, and scaled its local amplitude on the stellar surface with the function 
\begin{equation}
\label{eq:sc}
    S_c(\theta_{\rm T}) = 1 + \frac{2}{3}\cos\,\theta_{\rm T},
\end{equation} where $\theta_{\rm T}$ is the angle with respect to the tidal axis, such that the pulsation amplitude is up to five times larger in the hemisphere facing the primary than on the other side. Here, the value of the scaling parameter is not physically motivated, but merely set to $\frac{2}{3}$ for a conceptual illustration. The resulting amplitude and phase modulations of the simulated pulsation were then calculated using the methodology described in Section\,\ref{sec:methods}, and are illustrated in Figs. \ref{fig:obs-V456Cyg_model} and \ref{fig:V456Cyg_model}. In Fig.\,\ref{fig:V456Cyg_model}, we see the theoretical, high-resolution tidal perturbations, while in Fig.\,\ref{fig:obs-V456Cyg_model} the resolution has been downgraded to match the TESS data of V456\,Cyg. Given that this is just a toy model, the similarities between the simulations and the observations are striking. The spatial filtering signatures during the eclipses are well reproduced, and we see a similar relation between the amplitude and phase modulations as in our observations during the out-of-eclipse orbital phases. This indicates that the dominant tidal distortion effect on the g~modes is to amplify them on part of the tidally deformed stellar surface, while the observed pulsation phase modulations are at least partially caused by geometric effects of the pulsation mode visibility.

We note that despite this insight, our simulation does yet not provide any information on the physical processes that cause such distortion of the pulsation amplitude. The ellipsoidal variability of the pulsating star is likely a contributing factor, but as stated in Sect.\,\ref{sec:discussion}, the observed ellipsoidal variability and the amplitude modulations are not perfectly correlated, indicating that other aspects also contribute to the amplifications of the g~modes. Furthermore, our stars show different modulation of the mode amplitudes, which peak at or around orbital phases of 0.25 and 0.75 in KIC\,3228863, KIC\,3341457, KIC\,4947528, and KIC\,9108579 (Figs. \ref{fig:kic03228863_allpuls}, \ref{fig:kic03341457_allpuls}, \ref{fig:kic04947528_allpuls}, \ref{fig:kic09108579_allpuls}), at an orbital phase of 0 in KIC\,12785282 (Fig. \ref{fig:kic12785282_allpuls}), and at an orbital phase of 0.5 in V456\,Cyg (Fig. \ref{fig:obs-V456Cyg_model}). The functional form of the amplitude modulation will thus be different from equation \ref{eq:sc} for each case.

Unfortunately, the theoretical frameworks that have been presented in the literature do not suffice to describe these effects. None of them, including the study by \citet{Fuller2020}, have fully accounted for the effects of the Coriolis force, which is one of the two dominant forces known to affect g-mode pulsations \citep[e.g.][]{Mathis2009,Bouabid2013}. \citet{Fuller2020} modelled tidally tilted p-modes assuming they were aligned with the tidal axis, under the justification that tidal distortion was more important than the Coriolis and centrifugal forces. However, low-frequency gravity modes are more sensitive to Coriolis forces and are likely to remain aligned with the rotation axis, as observed in our sample because the mode period spacings closely follow the predictions of the TAR.

One possible mechanism to explain the amplitude variations is a ``tidal amplification" similar to that discussed in \cite{Fuller2020}. In a tidally distorted star, the internal sound speed and Brunt-V\"ais\"al\"a frequency profiles will not be spherically symmetric. This may change the geometry of the g-mode cavity, allowing modes to propagate closer to the surface and produce larger flux perturbations on different sides of the star. In the future, it will be necessary to develop a new tidal coupling framework to model mode amplitude variations and determine the true physical causes of the observed tidal perturbations.

\section{Rotational-modulation-like signals}
\label{sec:obs-rotmod}
In addition to the tidal g-mode perturbations, we also analysed the rotational-modulation-like variability of our five selected targets. While there are clear differences between the different binaries, it is often coherent over timescales of hundreds of days, and can provide us with additional information on the stellar properties. A proper understanding of the observed rotational-modulation-like signatures requires detailed modelling of our targets and lies outside of the scope of our current project. However, to facilitate future follow-up studies, we provide both observational characterisations and a qualitative discussion of the properties of these signals.

The rotational-modulation-like signatures have multiple possible physical origins, which are not always fully understood. In low-mass stars ($M \lesssim 1.3\,M_\odot$) rotational modulation is often a signature of spots on the stellar surface, caused by dynamic magnetic fields \citep[e.g.][]{Strassmeier2009,Chowdhury2018}, while in higher-mass stars ($M \gtrsim 2\,M_\odot$, with spectral types O, B, and A) fossil magnetic fields may lead to chemical abundance spots, though this does not explain all observations \citep[e.g.][and references therein]{David-Uraz2019b}. By contrast, it has also been proposed that in a sufficiently fast-rotating star with a convective core, overstable convective modes in the core can resonantly excite g-mode pulsations in the radiative envelope if it is rotating slightly slower than the core. The signal of such g~modes is then expected to resemble rotational spot modulation \citep[e.g.][]{Lee2020,Lee2021}. However, the rotational-modulation-like signal observed for our targets is much larger than what has been detected for (presumably) non-magnetic single stars \citep[e.g.][]{Li2020}, indicating that it is related to the binarity. Indeed, close binaries often have tertiary companions \citep[e.g.][]{Tokovinin2006}, which can induce long-term variations in the observed orbit \citep[e.g.][]{Toonen2016}. Additionally, the tidal deformation of stars in close binaries causes temperature, pressure, and density variations close to the stellar surface \citep[e.g.][]{Hilditch2001}, which can in turn induce circulation and flow processes that are visible on the stellar surface \citep[e.g.][]{Tassoul1982c,Tassoul1982b}. Over the last decade, these mechanisms have been studied extensively for the atmospheres of tidally locked exoplanets \citep[e.g.][]{ShowmanPolvani2011}. Here the atmospheric temperature variations were found to trigger standing equatorial Kelvin and Rossby waves, which can be tilted in the direction of the rotation by the balance between the pressure gradient, Coriolis, and drag forces in these targets \citep{ShowmanPolvani2011}.

\begin{figure}
    \centering
    \includegraphics[width=\columnwidth]{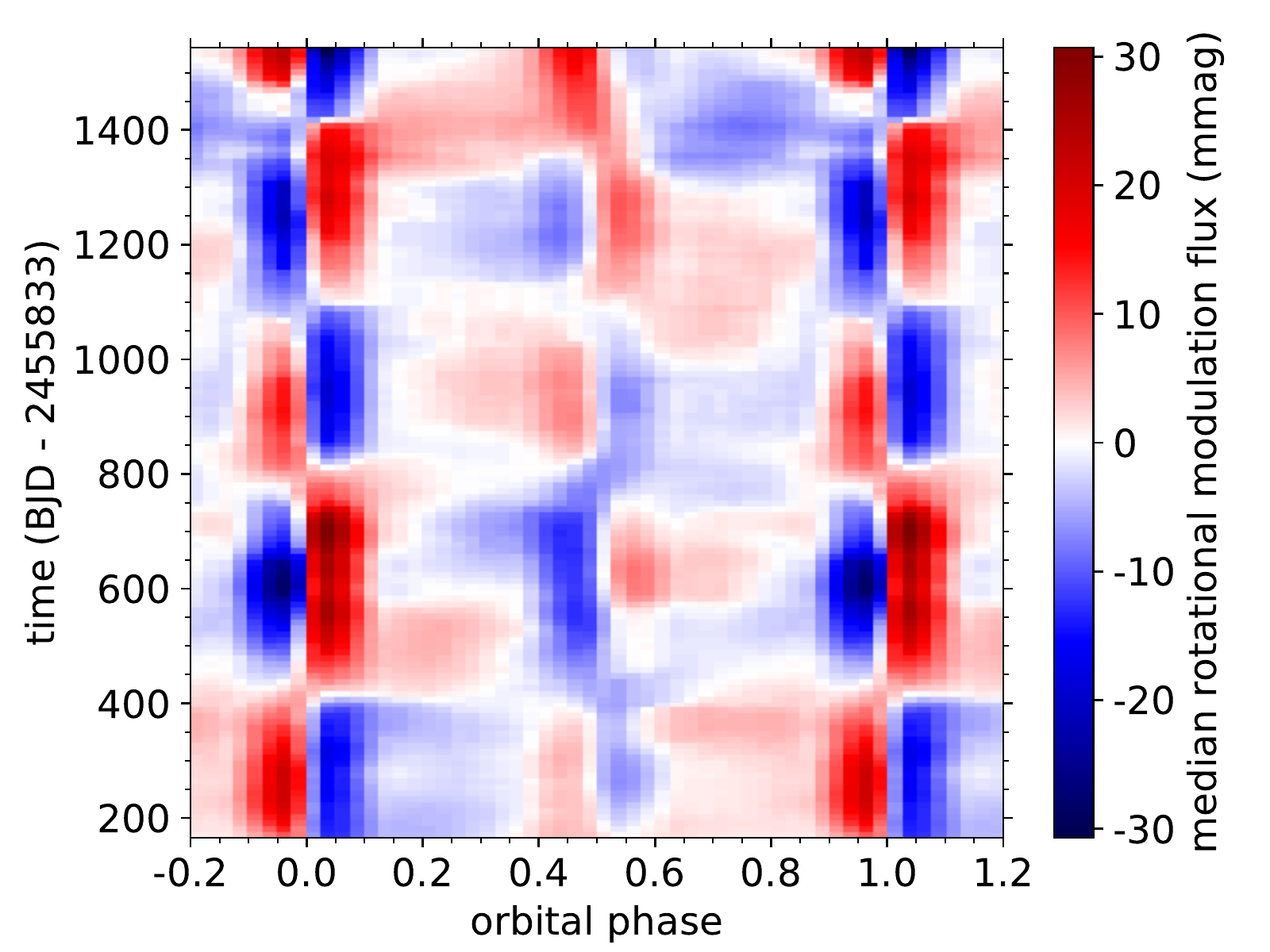}
    \caption{Time and orbital-phase dependence of the rotational-modulation-like signal of KIC\,3228863, calculated using 100-d sections of the light curve. The signal is dominated by the light-travel-time effect on the eclipses, caused by the presence of the tertiary component, as reported by \citet{Lee2014_V404Lyr,Lee2020_V404Lyr}.}
    \label{fig:kic03228863_rotmod}
\end{figure}

Our first target, KIC\,3228863, is the only one for which we could clearly determine the physical cause of the observed rotational-modulation-like signal. As shown in Fig.\,\ref{fig:kic03228863_rotmod}, the dependence of the signal as a function of time and orbital phase revealed that this variability is predominantly caused by timing variations in the eclipses. The measured periodicity of these variations are $\sim$643\,d, in agreement with the orbital period measured by \citet{Lee2020_V404Lyr} for the known tertiary component of this system ($P_{\rm orb} = 642 \pm 3$\,d).

\begin{figure}
    \centering
    \includegraphics[width=\columnwidth]{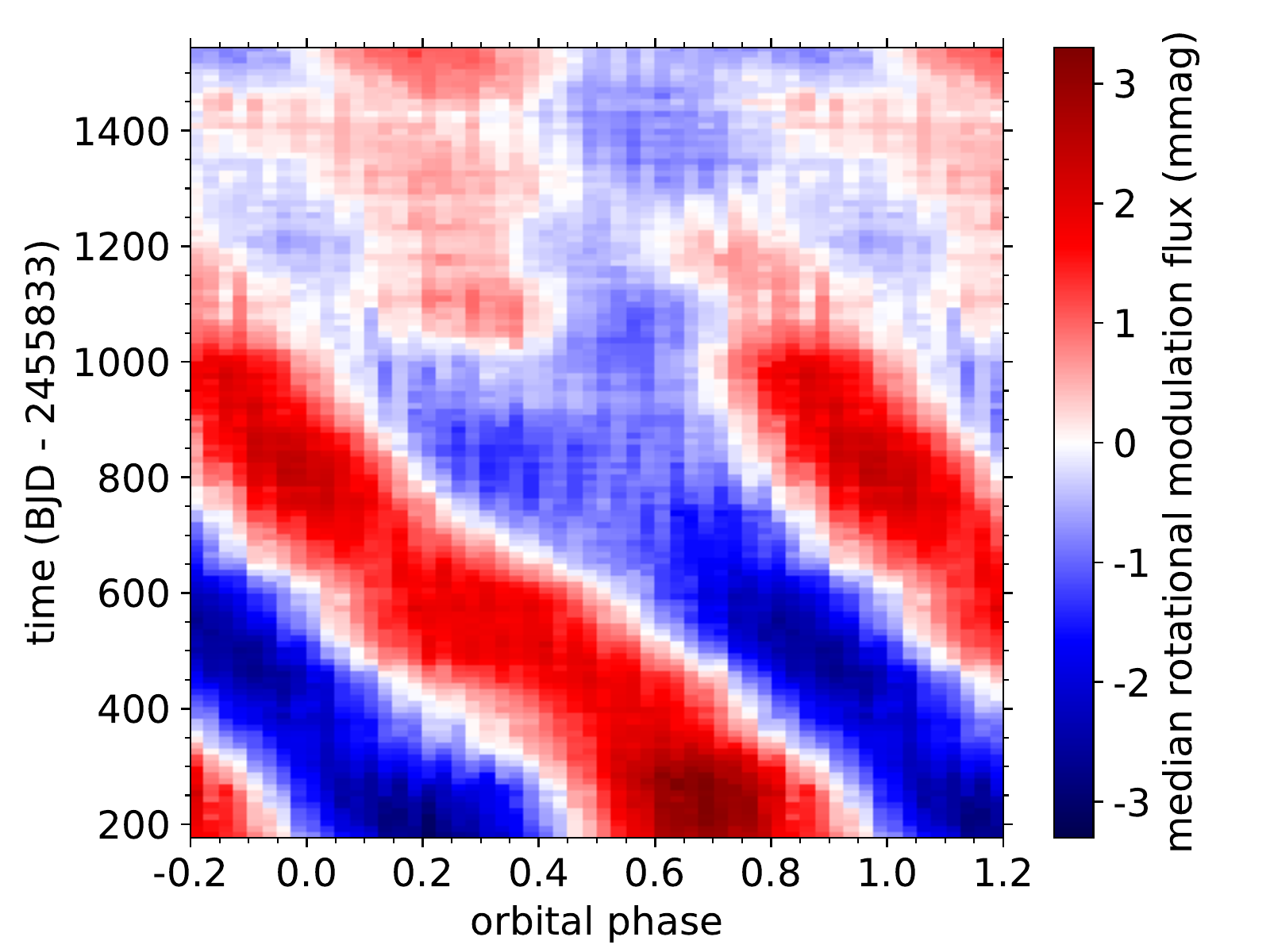}
    \caption{Time and orbital-phase dependence of the rotational-modulation-like signal of KIC\,3341457, calculated using 100-d sections of the light curve.}
    \label{fig:kic03341457_rotmod}
\end{figure}

\begin{figure}
    \centering
    \includegraphics[width=\columnwidth]{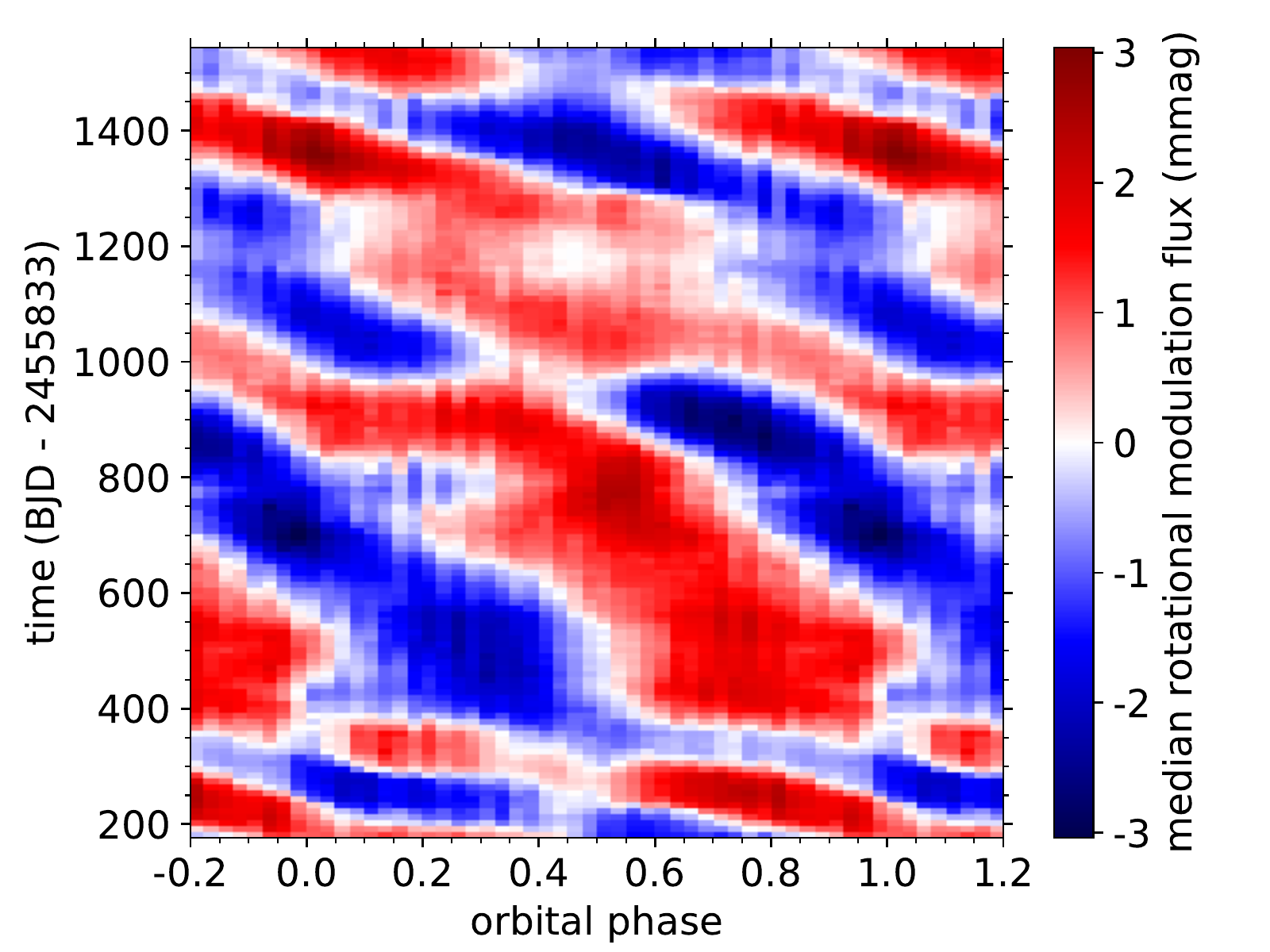}
    \caption{Time and orbital-phase dependence of the rotational-modulation-like signal of KIC\,4947528, calculated using 100-d sections of the light curve.}
    \label{fig:kic04947528_rotmod}
\end{figure}

\begin{figure}
    \centering
    \includegraphics[width=\columnwidth]{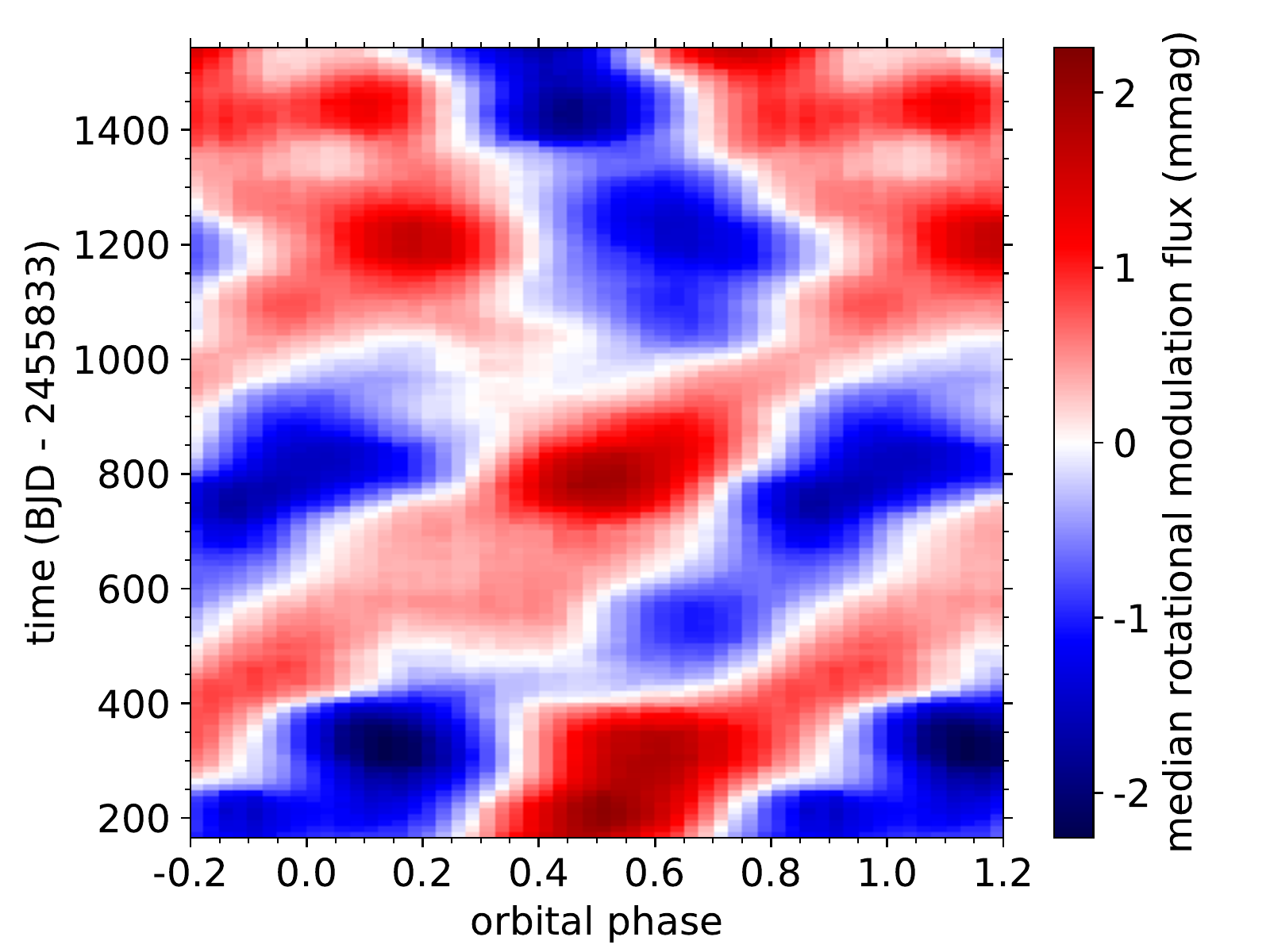}
    \caption{Time and orbital-phase dependence of the rotational-modulation-like signal of KIC\,9108579, calculated using 100-d sections of the light curve.}
    \label{fig:kic09108579_rotmod}
\end{figure}

By comparison, the physical origins of the rotational-modulation-like variability of KIC\,3341457, KIC\,4947528, and KIC\,9108579, as shown in Figs.\,\ref{fig:kic03341457_rotmod} to \ref{fig:kic09108579_rotmod}, could not be definitively determined, though they appear to be most consistent with large-scale temperature spots and associated flow patterns on the stellar surfaces. As shown by \citet{ShowmanPolvani2011} for the case of tidally locked exoplanets, thermal forcing can induce large-scale, standing Kelvin and Rossby waves. The toroidal velocity fields of these waves can then be tilted by the combined pressure gradient, Coriolis, and drag forces, which can in turn lead to an equatorial superinertial flow. We hypothesise that if a similar mechanism is present in our close binaries, this could explain the time dependence of the observed rotational-modulation-like signal of these three targets. For example, in the case of KIC\,3341457 it would correspond to an observed equatorial flow frequency of $\sim$1.89444\,$\rm d^{-1}$, which is slightly higher than the measured orbital frequency (1.8932976(5)\,$\rm d^{-1}$) even though the measured average near-core rotation rate is lower (1.865(1)\,$\rm d^{-1}$). Moreover, such a mechanical excitation of r~modes has also been hypothesised to occur in the so-called hump-and-spike stars \citep{Saio2018,Saio2022}. However, we cannot be certain; the observed rotational-modulation-like signal could also be caused by a different mechanism, such as overstable convective modes in the core \citep{Lee2020,Lee2021}. It is also possible that the rotational-modulation-like variability originates from the non-pulsating companion. Unfortunately, we currently do not have sufficient information to provide a definite answer. Time series of spectroscopic observations and more detailed theoretical modelling of these targets are required to evaluate this.

\begin{figure}
    \centering
    \includegraphics[width=\columnwidth]{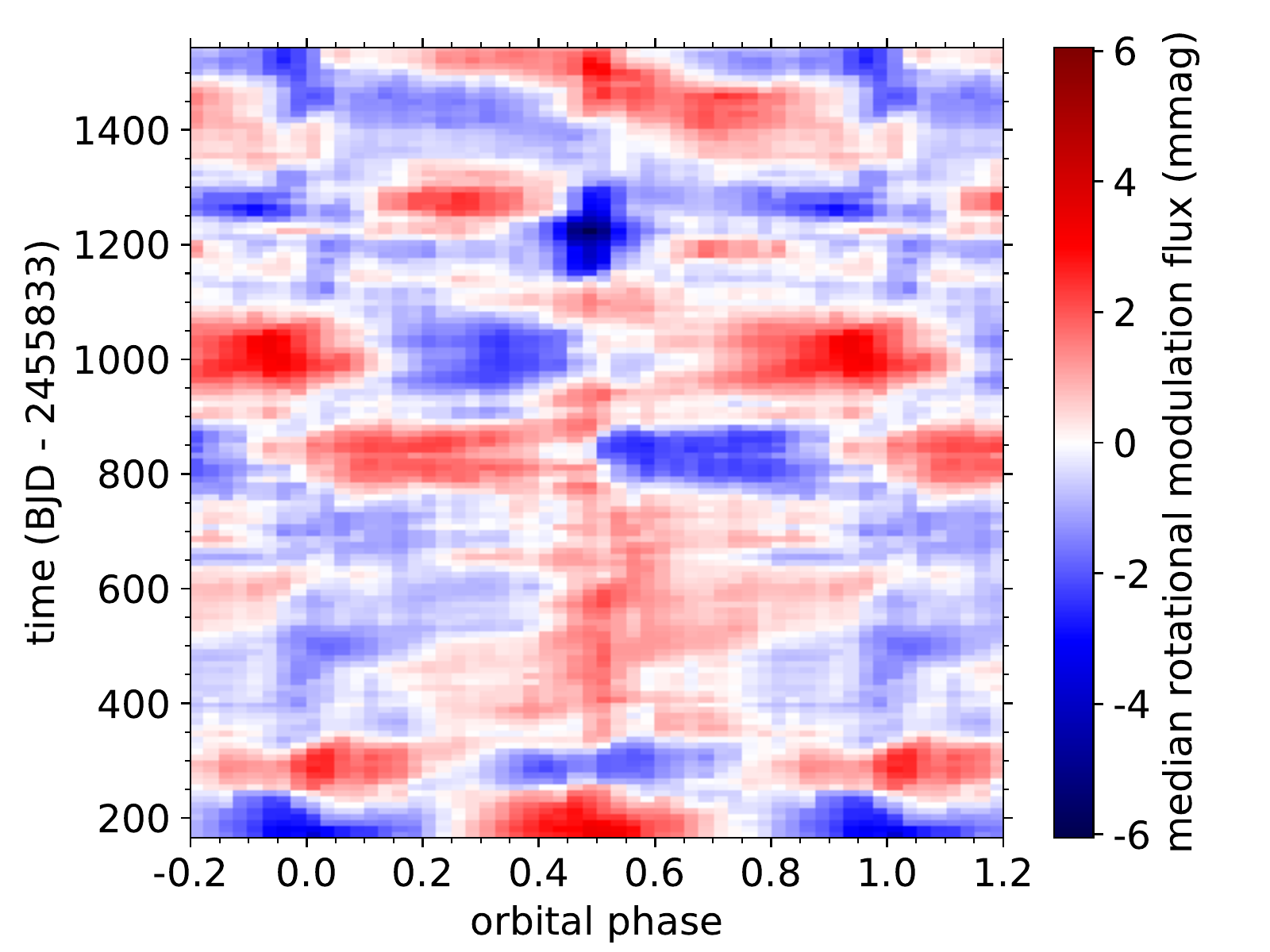}
    \caption{Time and orbital-phase dependence of the rotational-modulation-like signal of KIC\,12785282, calculated using 100-d sections of the light curve.}
    \label{fig:kic12785282_rotmod}
\end{figure}

The rotational-modulation-like signal of our last target, KIC\,12785282, differs significantly from our observations for the other stars, as shown in Fig.\,\ref{fig:kic12785282_rotmod}. It is less stable, and there are no clear long-term trends present. Furthermore, there is no clear regular time dependence during the eclipses, as observed for KIC\,3228863, which was caused by the presence of the tertiary component. However, it is still possible that such signal is present on timescales shorter than 100\,d. We used 100-d windows in our analysis to ensure numerical stability, but the downside is that we could not detect any shorter-term variability. As a result, we could not determine a (clear) physical cause for the observed rotational-modulation-like signal of this star either.

Finally, we note that in addition to our five selected targets, the rotational-modulation-like signal is also present in some, but not all, of the other close binaries in the sample from \citet{Li2020_bin}. Examples of these targets include KIC\,6048106, KIC\,6206751, KIC\,7385472, KIC\,8197406 and KIC\,8330092. So while we cannot definitively identify the true physical cause of the rotational-modulation-like signal, we can conclude that it is not directly related to the tidal perturbation of g-mode pulsations. And given that it appears to only be present in some close binaries, but not in wide binaries or single stars \citep[e.g.][]{Li2020}, it seems likely that it is causally related to the properties of close binaries. However, a detailed analysis of these targets is required to resolve this issue.

\section{Conclusions}
\label{sec:conclusions}
We have conducted a detailed asteroseismic analysis of the 35 $\gamma$\,Dor stars in binary systems with g-mode period-spacing patterns, presented by \citet{Li2020_bin}, and detected tidal perturbation of the pulsations in the five shortest-period binaries of their sample, which are circularised and have orbital frequencies higher than or of the order of $1\,\rm d^{-1}$. Combined with the previously discovered binary with tidally perturbed $\gamma$\,Dor type pulsations (V456\,Cyg; \citealt{VanReeth2022a}), this allowed us to conduct a first ensemble characterisation of this phenomenon. The tidally perturbed g-mode pulsations that were previously reported for the SPB pulsator $\pi^5$\,Ori \citep{Jerzykiewicz2020}, were not included in this study to avoid a potential bias. In addition to this star belonging to a different pulsator class, its analysis was based on data from the BRITE constellation \citep{Weiss2014} rather than {\em Kepler} or TESS.

First, we have found that the relative amplitude and phase modulations of tidally perturbed g~modes within a given star depend strongly on the mode geometry. For pulsations with the same mode identification, they are almost identical. In other words, they are observed to be independent of the intrinsic g-mode pulsation amplitudes, radial orders, and phases. Moreover, based on a comparison of the available {\em Kepler} and TESS photometry and within the limits of the available data, the relative modulations are found to be independent of the photometric passband in which they are observed. By contrast, the tidal modulations of g-mode pulsations with different mode geometries $(k,m)$ exhibit clear differences. As these pulsations have different mode cavities, velocity, and displacement fields, they have different sensitivities to the tidal distortion of the pulsating star and to the Coriolis and centrifugal forces in the binary system.

Second, the dominant observed characteristic of tidally perturbed g-mode pulsations is a distortion of the pulsation amplitude on the stellar surface, which is related to the tidal deformation of the pulsating star. This affects both the observed pulsation amplitude and phase modulations, and results in similar orbital-phase-dependent signatures as we see in the data of our selected sample. In most of our sample, the mode amplitudes are largest at or around orbital phases of 0.25 and 0.75 (when we are looking at the tidally distorted star side-on), but in a few cases the amplitude peaks at an orbital phase of 0.5 (when we are looking at the narrow side of the pulsating star that is facing the binary companion), and in other cases the amplitude modulation is asymmetric with orbital phase. Hence, the physical origins of these amplitude distortions are not yet clear. This requires a detailed theoretical study, which given that we are dealing with sub-inertial g~modes should also account for the effects of the Coriolis acceleration on the pulsations. Such work lies outside of the scope of the presented observational analysis.

Interestingly, all five targets presented in this work also exhibit rotational-modulation-like signal which, except for KIC\,12785282, contains clear coherent variability as a function of time. For KIC\,3228863, this signal is a result of eclipse timing variations, caused by the known tertiary \citep{Lee2014_V404Lyr,Lee2020_V404Lyr} that orbits the observed close binary. By contrast, the true physical origins of the rotational-modulation-like variability are currently still unknown for the remaining three stars, and we hypothesise that they are caused by temperature variations and associated flow patterns on the stellar surface. Alternatively, they could be caused by overstable convective modes in the stellar core, belong to the non-pulsating binary component, or be related to other variations in the binary motion. However, in each of these cases, the observed rotational-modulation-like signal provides information on the stellar properties, complementing the information from the g-mode pulsations and the binarity. Hence, when the physical origins of these rotational-modulation-like signals are known, they might be used to help constrain the angular momentum transport processes in our five targets.

In conclusion, our results indicate that with a suitable theoretical framework the g-mode perturbations can be used to probe the deformation of the outer layers of the pulsating stars, where the tidal effects are dominant. Since g-mode periods mostly provide information on the deep interior stellar structure, tidally perturbed g-mode pulsations provide us with a unique opportunity to study the stellar structure from the near-core region to the surface. The five binaries presented in this work are excellent targets for such a study. Moreover, other sample targets may be as well. Our strict detection criteria for tidal perturbations were specifically chosen to ensure unambiguous detections. It is possible that other stars in our sample exhibit tidally perturbed g~modes that did not meet our detection criteria.

\begin{acknowledgements}
Combinations and distortions of different pulsation mode geometries can be complicated. TVR is very grateful to Marrick Braam, Sarah Gebruers, Mathias Michielsen, and Joey S. G. Mombarg for useful discussions that have helped him to make sense of everything, or at least get confused consistently. We also thank the anonymous referee for useful comments which have improved the quality of the manuscript.

TVR and DMB gratefully acknowledge funding from the research foundation Flanders (FWO) by means of junior and senior postdoctoral fellowships with grant agreements N${}^\circ$ 12ZB620N and 1286521N, respectively, and FWO long stay travel grants N${}^\circ$ V414021N and V411621N, respectively. JVB acknowledges receiving support from the Research Foundation Flanders (FWO) under grant agreement N${}^\circ$ V421221N. The research leading to these results also received partial funding from the KU Leuven Research Council (grant C16/18/005: PARADISE), and was supported in part by the National Science Foundation under Grant No. NSF PHY-1748958. CJ gratefully acknowledges support from the Netherlands Research School of Astronomy (NOVA).

This paper includes data collected with the {\em Kepler} and TESS missions, obtained from the MAST data archive at the Space Telescope Science Institute (STScI). Funding for the {\em Kepler} and TESS missions are provided by NASA's Science Mission Directorate and the NASA Explorer Program, respectively. We thank the whole teams for the development and operations of these missions. STScI is operated by the Association of Universities for Research in Astronomy, Inc., under NASA contract NAS 5-26555. Support to MAST for these data is provided by the NASA Office of Space Science via grant NAG5-7584 and by other grants and contracts. 

This research also made use of the SIMBAD database, operated at CDS, Strasbourg, France, the SAO/NASA Astrophysics Data System, and the VizieR catalogue access tool, CDS, Strasbourg, France. The data analysis was done using Astropy\footnote{\href{http://www.astropy.org}{http://www.astropy.org}} \citep[a community-developed core Python package for Astronomy;][]{astropy:2013,astropy:2018}, Astroquery\footnote{\href{https://astroquery.readthedocs.io/en/latest/}{https://astroquery.readthedocs.io/en/latest/}} \citep{astroquery}, Lightkurve\footnote{\href{https://docs.lightkurve.org/}{https://docs.lightkurve.org/}} \citep[a Python package for Kepler and TESS data analysis;][]{Lightkurve2018}, lmfit \citep{lmfit100}, Matplotlib \citep[the Python library for publication quality graphics;][]{Hunter2007}, Numpy \citep{numpy}, and Scipy \citep{scipy}.
\end{acknowledgements}

\bibliographystyle{aa}
\bibliography{Tidally-perturbed-g-modes}

\begin{appendix}

\section{Semi-analytical tidal perturbation model}
\label{sec:semi-analytical-perturbations}
Following the example by \citet{Jayaraman2022}, we used the pulsations that were prewhitened and presented in Section~\ref{subsec:pulsations} to build a semi-analytical model $M_j(t)$ of the photometric variability of each tidally perturbed pulsation with frequency $\nu_j$:
\begin{equation}
    M_j\left(t\right) = \sum_n a_{j,n}\sin\left(2\pi\left[\left(\nu_j + n\nu_{\rm orb}\right)\left(t - t_0\right) + \varphi_{j,n}\right]\right),
\end{equation}
with $n \in \mathbb{Z}$ indicating the different components of the $\nu_{\rm orb}$-spaced multiplet around $\nu_j$. This expression can be rewritten as
\begin{equation}
    M_j\left(t\right) = a_{\nu}\left(\phi_{\rm orb}\right)\sin\left(2\pi\left[\nu \left(t-t_0\right) + \varphi_{\nu}\left(\phi_{\rm orb}\right)\right]\right),
\end{equation}
where the orbital phase $\phi_{\rm orb} = \nu_{\rm orb}\left(t-t_0\right)$, and $a_{\nu,j}$ and $\varphi_{\nu,j}$ are the orbital-phase-dependent amplitude and phase of pulsation frequency $\nu_j$, respectively. These are then given by
\begin{equation}
    \begin{split}
    a_{\nu,j}\left(\phi_{\rm orb}\right) = &\left[\left(\sum_n a_{j,n}\sin\left(2\pi\left[n\phi_{\rm orb} + \varphi_{j,n}\right]\right)\right)^2\right.\\
                                         & \left.\qquad\qquad+ \left(\sum_n a_{j,n}\cos\left(2\pi\left[n\phi_{\rm orb} + \varphi_{j,n}\right]\right)\right)^2\right]^\frac{1}{2},\label{eq:ampmod}
    \end{split}
\end{equation}
and
\begin{equation}
    \begin{split}
    \varphi_{\nu,j}\left(\phi_{\rm orb}\right) = \frac{1}{2\pi}\arctan& \left[\frac{\sum_n a_{j,n}\sin\left(2\pi\left[n\phi_{\rm orb} + \varphi_{j,n}\right]\right)}{\sum_n a_{j,n}\cos\left(2\pi\left[n\phi_{\rm orb} + \varphi_{j,n}\right]\right)}\right].\label{eq:phasemod}
    \end{split}
\end{equation}
When $\nu_j + n\nu_{\rm orb} > 0$, the corresponding amplitude and phase values $a_{j,n}$ and $\varphi_{j,n}$ are simply those of the pulsations measured in Section\,\ref{subsec:pulsations} that fulfil the frequency requirements. However, when $\nu_j + n\nu_{\rm orb} < 0$, or in other words $\nu_j < |n\nu_{\rm orb}|$ with $n < 0$, the pulsation phase $\varphi_{j,n}$ used in Eqs. (\ref{eq:ampmod}) and (\ref{eq:phasemod}) is given by 
\begin{equation*}
    \varphi_{j,n} = \frac{1}{2} - \varphi_{i},
\end{equation*} 
where $\varphi_{i}$ is the phase value measured in the prewhitening in Sect.\,\ref{subsec:pulsations}. 


\newpage
\section{Detections of tidally perturbed g~modes}
\label{appendix:tidal-detections}
In this Appendix we have collected figures that illustrate the data analysis for the stars with detected tidally perturbed g-mode pulsations. For each star, we show {\em (i)} the detected period-spacing patterns in the Lomb-Scargle periodogram of the out-of-eclipse parts of the pulsation light curve, that is the residual {\em Kepler} light curve after removing the best-fitting harmonic model, {\em (ii)} the detected orbital-frequency-spaced multiplets for the independent g-mode pulsations, and {\em (iii)} the observed tidal perturbations of the dominant g~mode(s), as a function of the binary orbital phase. To illustrate these, we show both the semi-analytical model that was described in Appendix\,\ref{sec:semi-analytical-perturbations}, and the binned data analysis of the modulations of the dominant g~mode. Because the sine waves that were used to build the model, were determined from the out-of-eclipse data, there can be clear offsets between the model and the binned data during the eclipses (if present). These signatures are caused by the partial occultation of the pulsation mode geometry, also called eclipse mapping \citep{Reed2001,Reed2005,Johnston2022} or spatial filtering \citep{Gamarova2003,Rodriguez2004,Gamarova2005}, and indicate to which binary component the observed pulsations belong.

\begin{figure*}
    \centering
    \includegraphics[width=\textwidth]{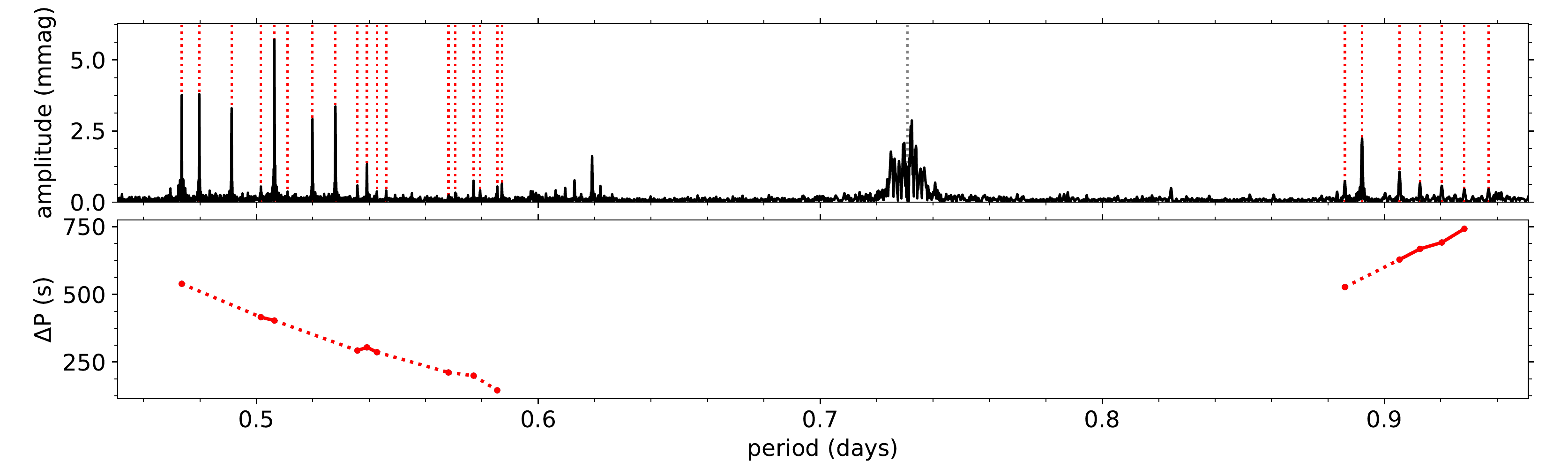}
    \caption{Detected period-spacing patterns of KIC\,3228863. {\em Top:} Lomb-Scargle periodogram (black) with the g~modes (red dotted lines) that form a $(k,m) = (0,1)$~pattern and a (-2,-1) pattern on the left- and right-hand side, respectively. The grey dotted line indicates an orbital harmonic. {\em Bottom:} Period spacing as a function of pulsation period for the detected g-mode patterns. Dotted lines indicate gaps in the patterns.\label{fig:kic03228863_period-spacings}}
\end{figure*}

\begin{figure*}
    \centering
    \includegraphics[width=\textwidth]{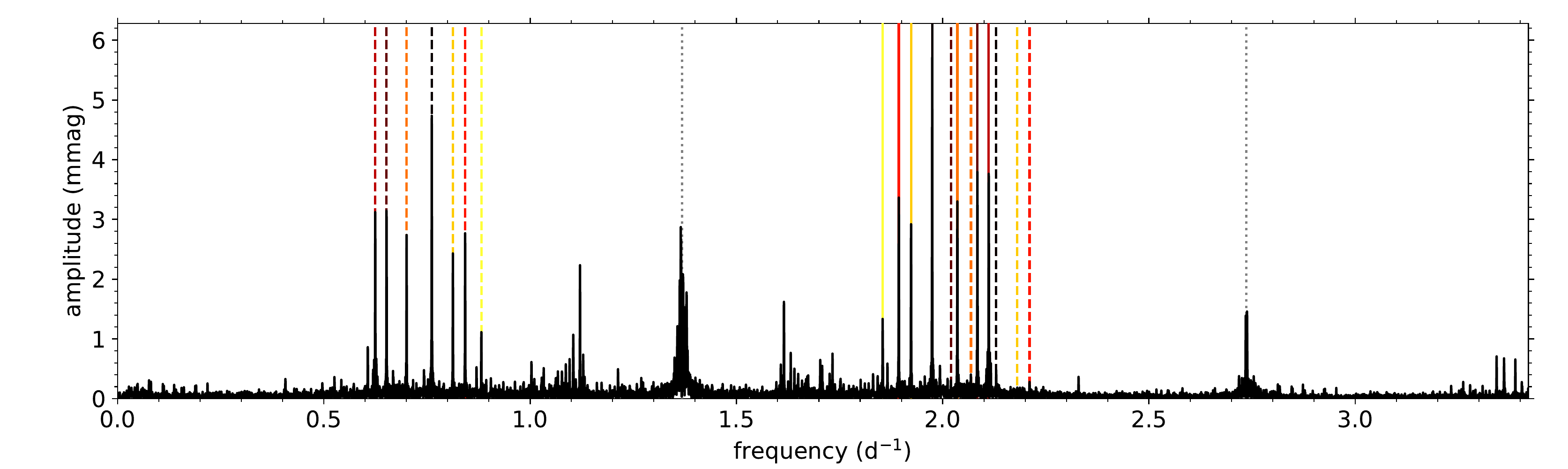}
    \caption{Orbital-frequency-spaced multiplets of KIC\,3228863. Lomb-Scargle periodogram (black) with the independent g-mode pulsations (full lines) and their multiplet components (dashed lines) shown in different matching colours. The grey dotted lines indicate orbital harmonics. \label{fig:kic03228863_fourier}}
\end{figure*}

\begin{figure}
    \centering
    \includegraphics[width=\columnwidth]{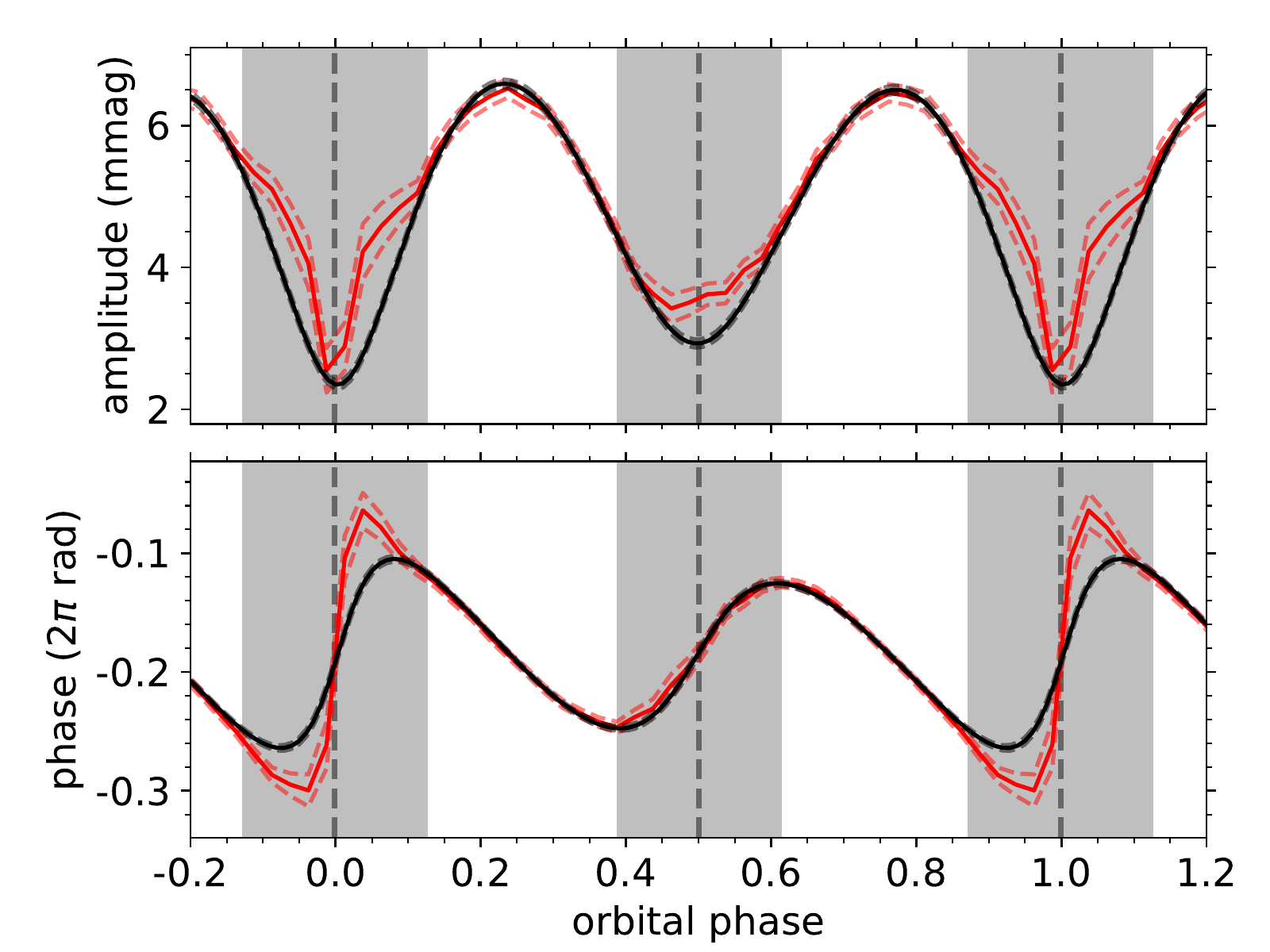}
    \caption{Tidal perturbation of the dominant $(k,m) = (0,1)$ g~mode of KIC\,3228863, with $\nu = 1.974605(3)\,\rm d^{-1}$. {\em Top:} pulsation amplitude modulations as a function of the binary orbital phase, calculated numerically within separate orbital phase bins (red) and reconstructed using prewhitened sine waves (black), with their $1-\sigma$ uncertainties (thin dashed lines). The binary eclipses are marked in grey. {\em Bottom:} pulsation phase modulations as a function of the binary orbital phase, calculated numerically from binned data (red) and reconstructed using prewhitened sine waves (black).\label{fig:kic03228863_bestpuls}}
\end{figure}


\begin{figure*}
    \centering
    \includegraphics[width=\textwidth]{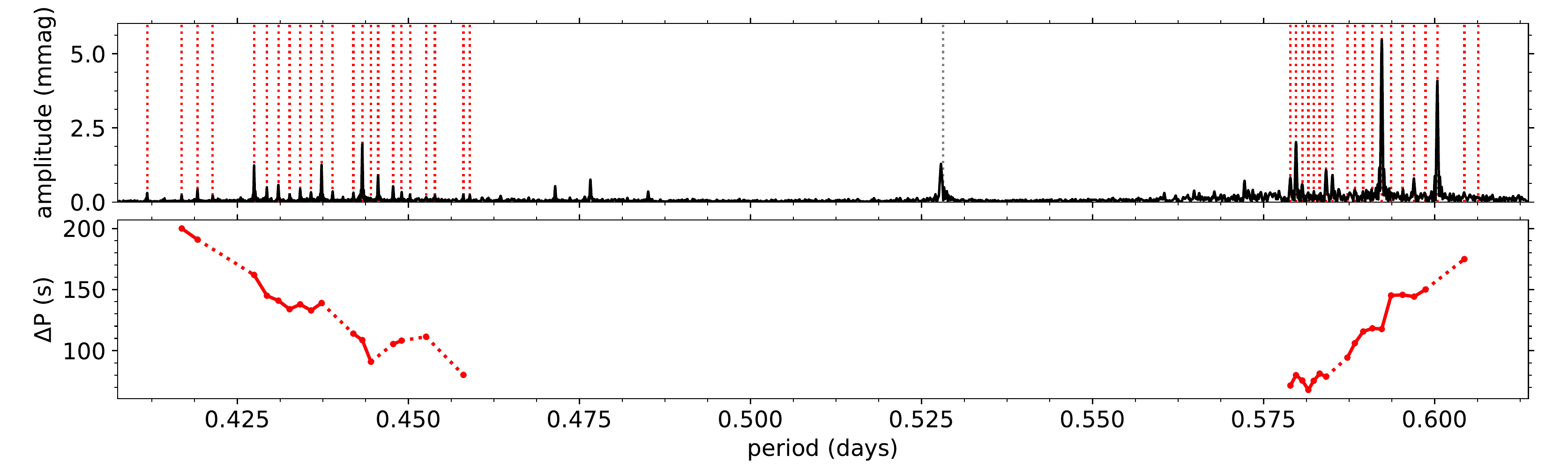}
    \caption{Detected period-spacing patterns of KIC\,3341457. {\em Top:} Lomb-Scargle periodogram (black) with the g~modes (red dotted lines) that form a $(k,m) = (0,1)$~pattern and a (-2,-1)~pattern on the left- and right-hand side, respectively. The grey dotted line indicates an orbital harmonic. {\em Bottom:} Period spacing as a function of pulsation period for the detected g-mode patterns. Dotted lines indicate gaps in the patterns.\label{fig:kic03341457_period-spacings}}
\end{figure*}

\begin{figure*}
    \centering
    \includegraphics[width=\textwidth]{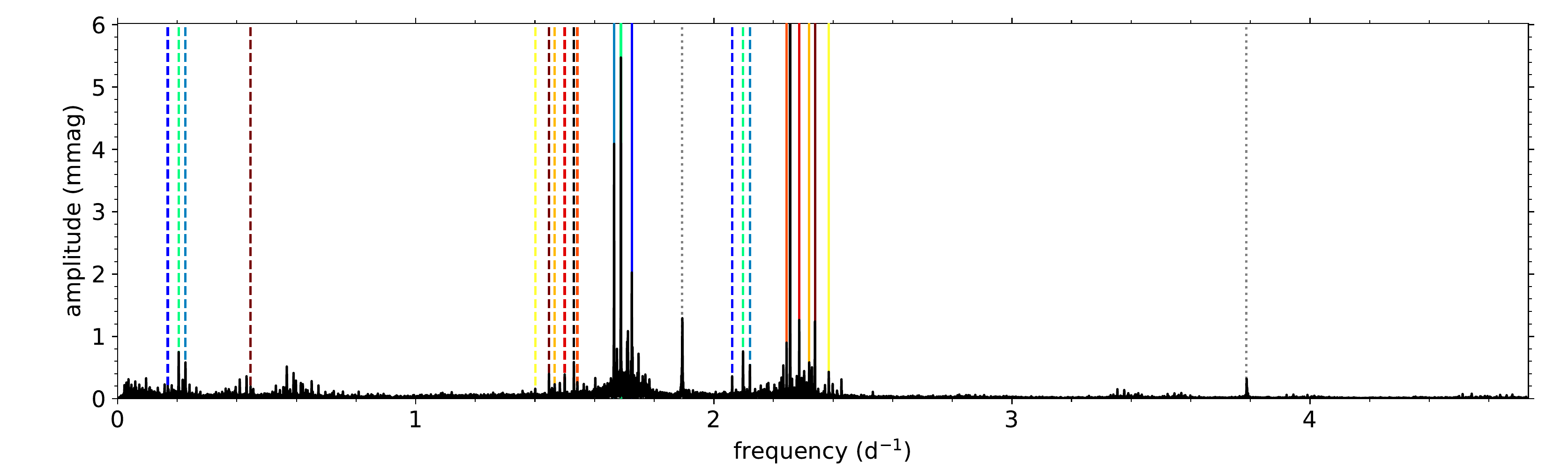}
    \caption{Orbital-frequency-spaced multiplets of KIC\,3341457. Lomb-Scargle periodogram (black) with the independent g-mode pulsations (full lines) and their multiplet components (dashed lines) shown in different matching colours. Pulsations with $(k,m) = (0,1)$ are marked in dark red tot yellow, while (-2,-1)~modes are shown in shades of blue to green. The grey dotted lines indicate orbital harmonics. \label{fig:kic03341457_fourier}}
\end{figure*}

\begin{figure}
    \centering
    \includegraphics[width=\columnwidth]{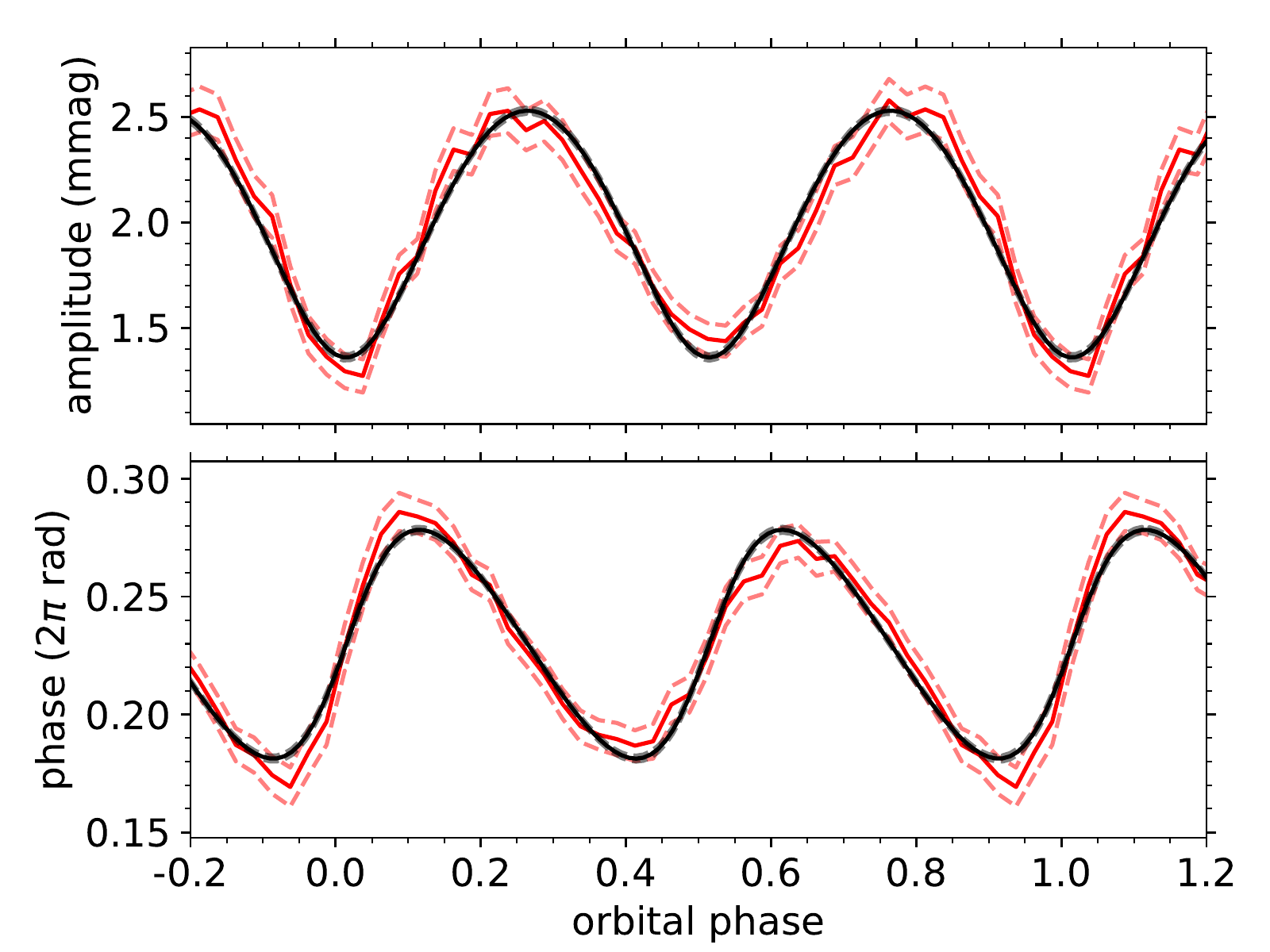}
    \caption{Tidal perturbation of the dominant $(k,m) = (0,1)$ g~mode of KIC\,3341457, with $\nu = 2.255988(2)\,\rm d^{-1}$, similar to Fig.\,\ref{fig:kic03228863_bestpuls}.\label{fig:kic03341457_bestgpuls}}
\end{figure}
\begin{figure}
    \centering
    \includegraphics[width=\columnwidth]{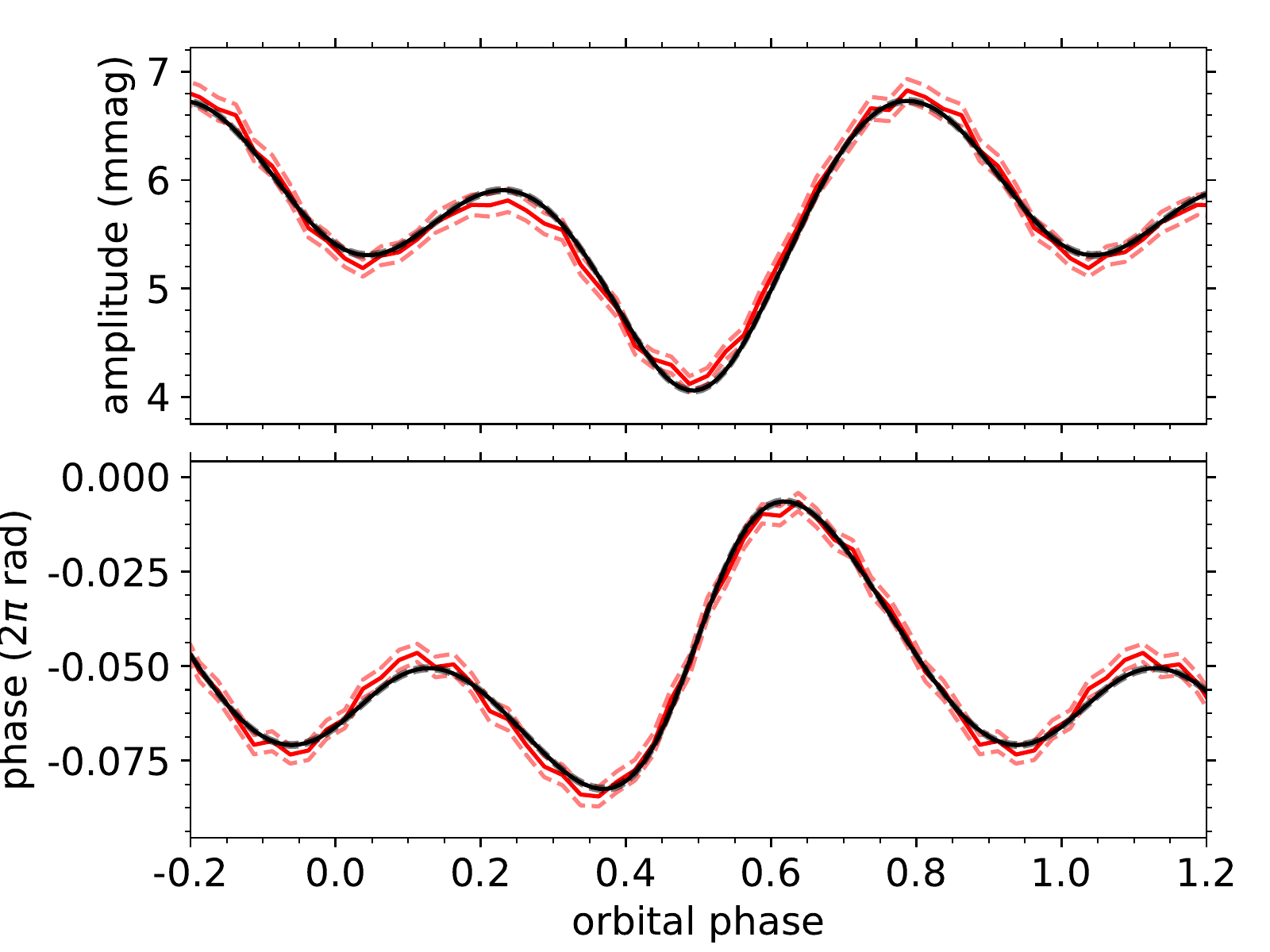}
    \caption{Tidal perturbation of the dominant $(k,m) = (-2,-1)$ g~mode of KIC\,3341457, with $\nu = 1.6884191(7)\,\rm d^{-1}$, similar to Fig.\,\ref{fig:kic03228863_bestpuls}.\label{fig:kic03341457_bestrpuls}}
\end{figure}


\begin{figure*}
    \centering
    \includegraphics[width=\textwidth]{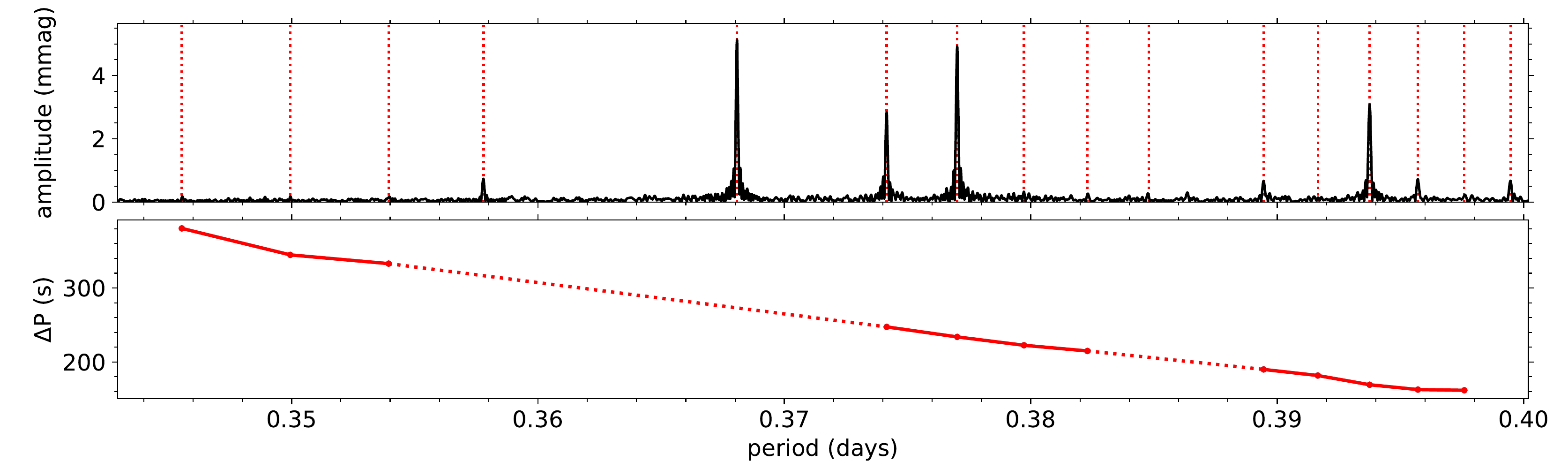}
    \caption{Detected period-spacing patterns of KIC\,4947528. {\em Top:} Lomb-Scargle periodogram (black) with the g~modes (red dotted lines) that form a $(k,m) = (0,1)$~pattern. {\em Bottom:} Period spacing as a function of pulsation period for the detected g-mode pattern, with dotted lines indicating gaps.\label{fig:kic04947528_period-spacings}}
\end{figure*}

\begin{figure*}
    \centering
    \includegraphics[width=\textwidth]{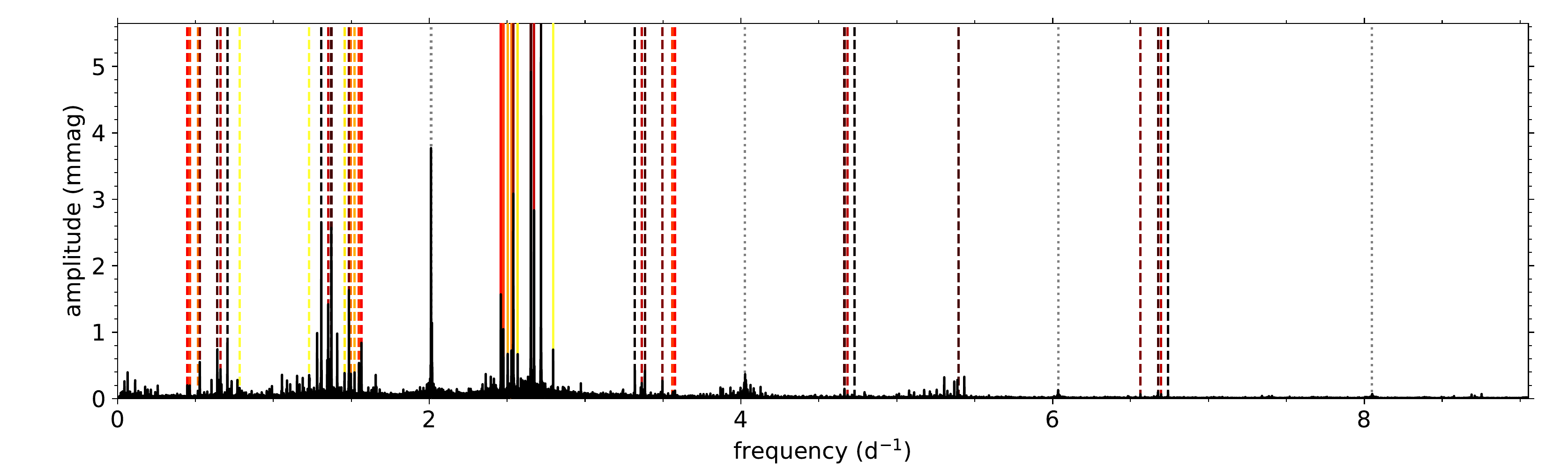}
    \caption{Orbital-frequency-spaced multiplets of KIC\,4947528. Lomb-Scargle periodogram (black) with the independent g-mode pulsations (full lines) and their multiplet components (dashed lines) shown in different matching colours. The grey dotted lines indicate orbital harmonics. \label{fig:kic04947528_fourier}}
\end{figure*}

\begin{figure}[h!]
    \centering
    \includegraphics[width=\columnwidth]{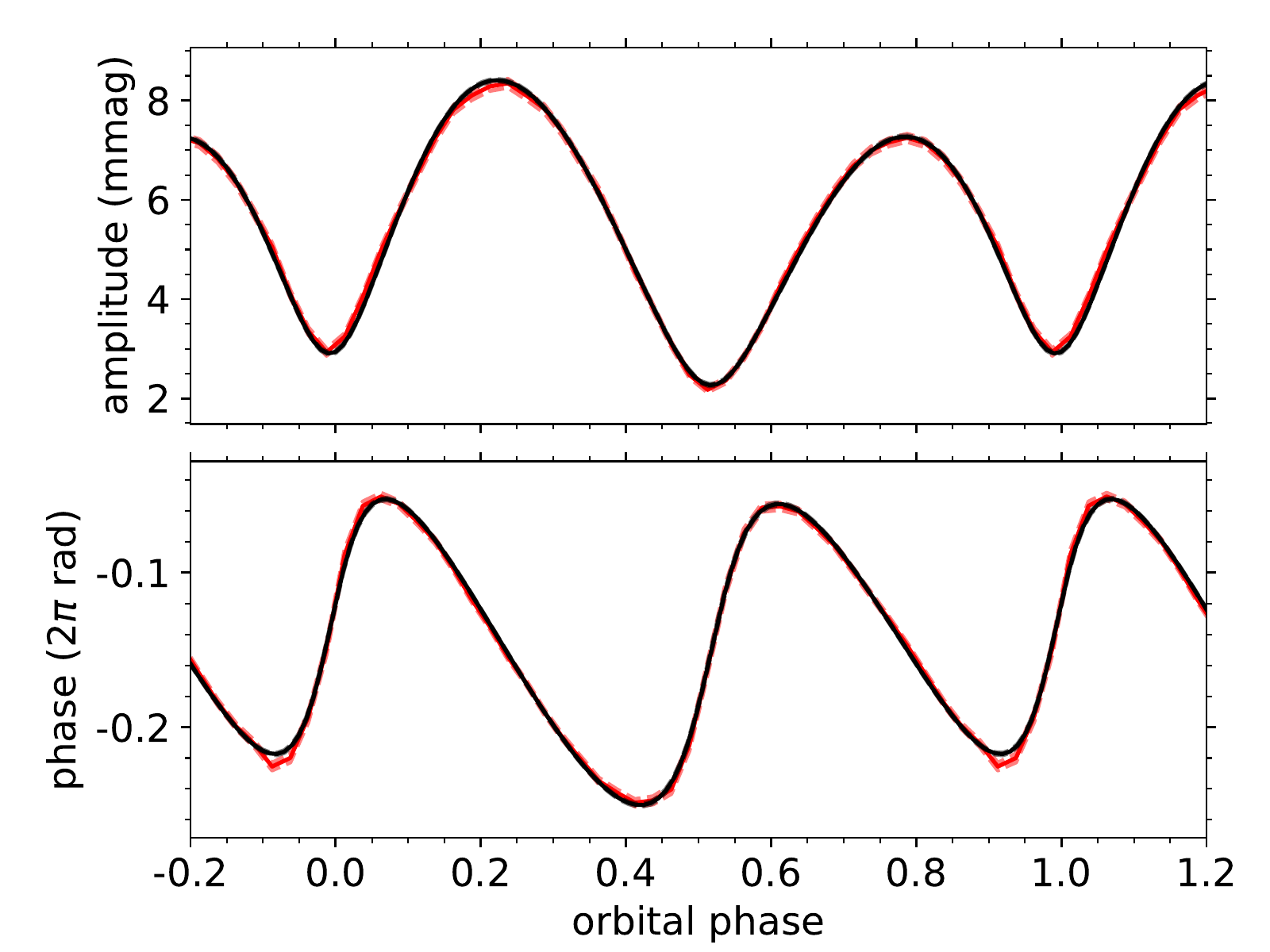}
    \caption{Tidal perturbation of the dominant $(k,m) = (0,1)$ g~mode of KIC\,4947528, with $\nu = 2.7168248(7)\,\rm d^{-1}$, similar to Fig.\,\ref{fig:kic03228863_bestpuls}.\label{fig:kic04947528_bestpuls}}
\end{figure}


\begin{figure*}
    \centering
    \includegraphics[width=\textwidth]{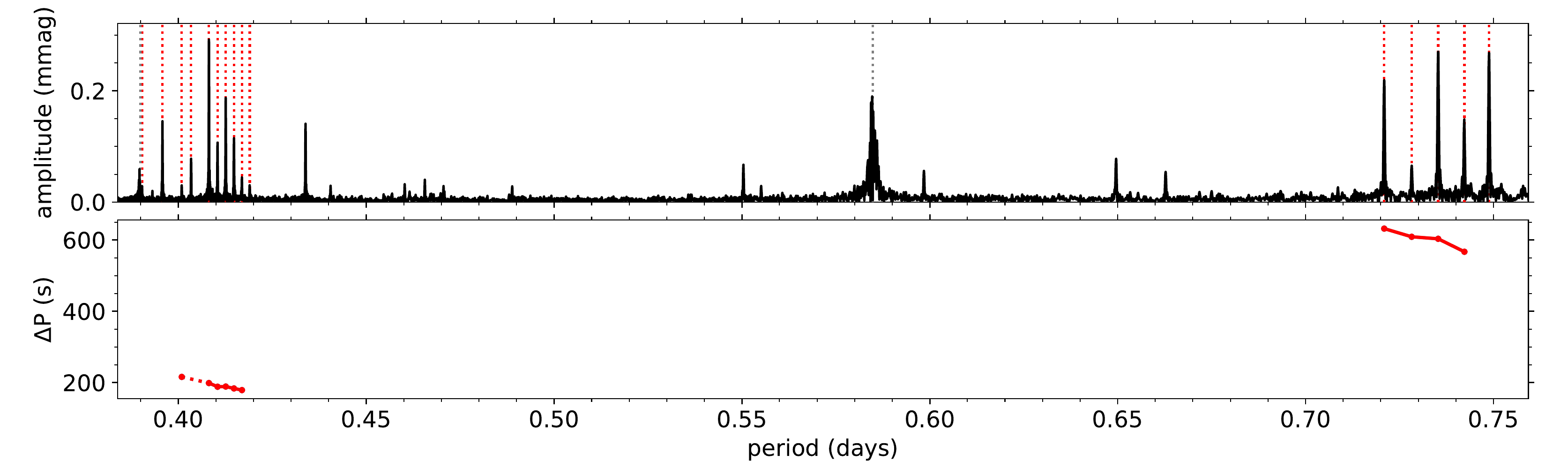}
    \caption{Detected period-spacing patterns of KIC\,9108579. {\em Top:} Lomb-Scargle periodogram (black) with the g~modes (red dotted lines) that form a $(k,m) = (0,2)$~pattern and a (0,1) pattern on the left- and right-hand side, respectively. The grey dotted line indicates an orbital harmonic. {\em Bottom:} Period spacing as a function of pulsation period for the detected g-mode patterns. Dotted lines indicate gaps in the patterns.\label{fig:kic09108579_period-spacings}}
\end{figure*}

\begin{figure*}
    \centering
    \includegraphics[width=\textwidth]{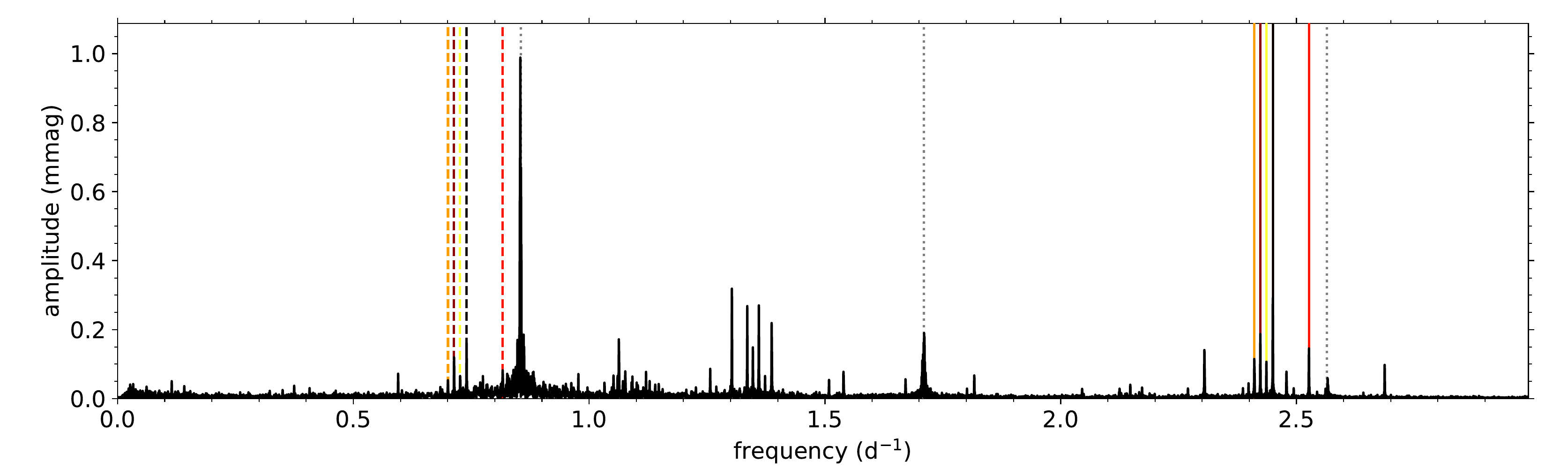}
    \caption{Orbital-frequency-spaced multiplets of KIC\,9108579. Lomb-Scargle periodogram (black) with the independent g-mode pulsations (full lines) and their multiplet components (dashed lines) shown in different matching colours. The grey dotted lines indicate orbital harmonics. \label{fig:kic09108579_fourier}}
\end{figure*}

\begin{figure}
    \centering
    \includegraphics[width=\columnwidth]{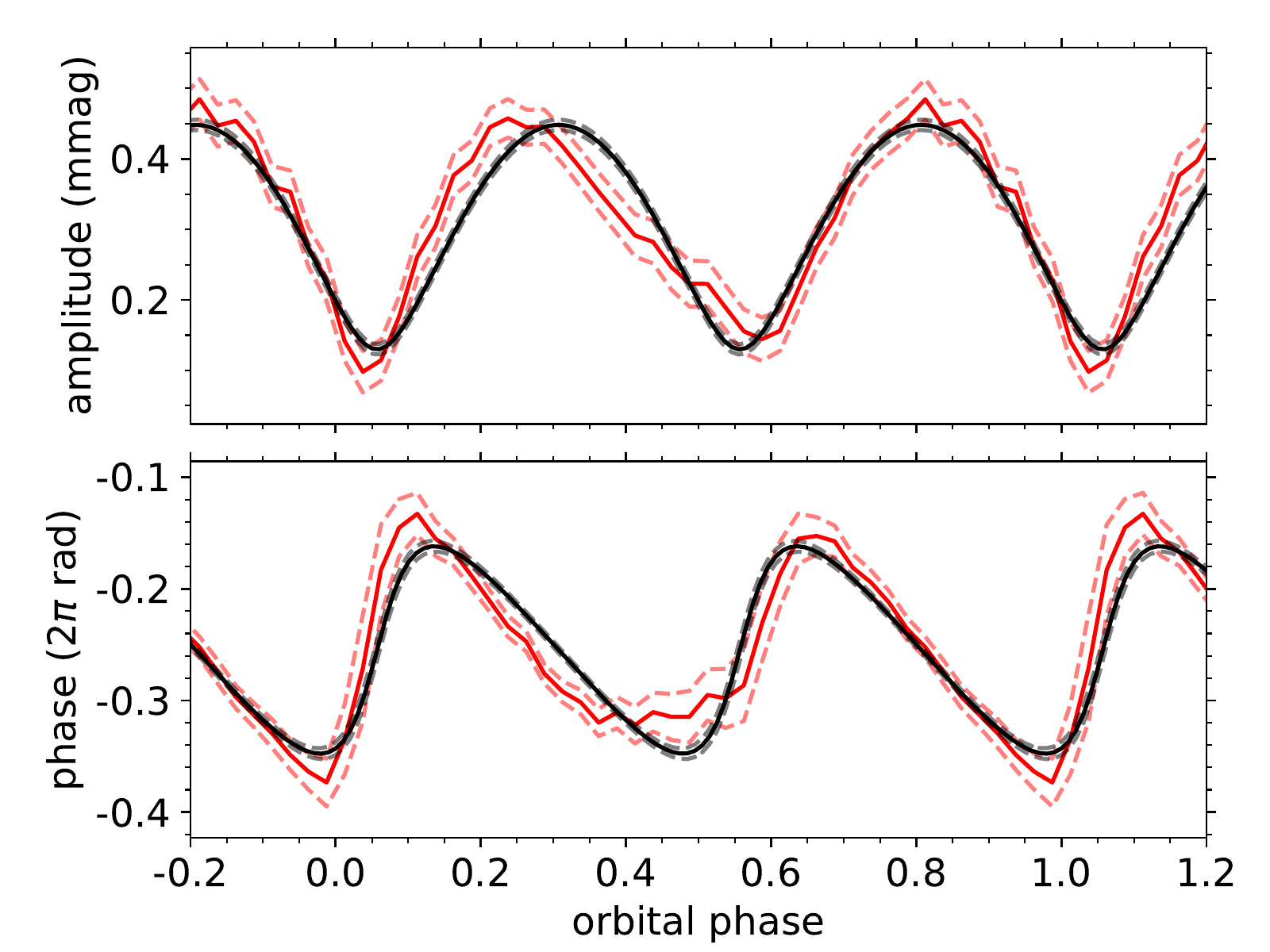}
    \caption{Tidal perturbation of the dominant $(k,m) = (0,2)$ g~mode of KIC\,9108579, with $\nu = 2.450087(7)\,\rm d^{-1}$, similar to Fig.\,\ref{fig:kic03228863_bestpuls}.\label{fig:kic09108579_bestpuls}}
\end{figure}


\begin{figure*}
    \centering
    \includegraphics[width=\textwidth]{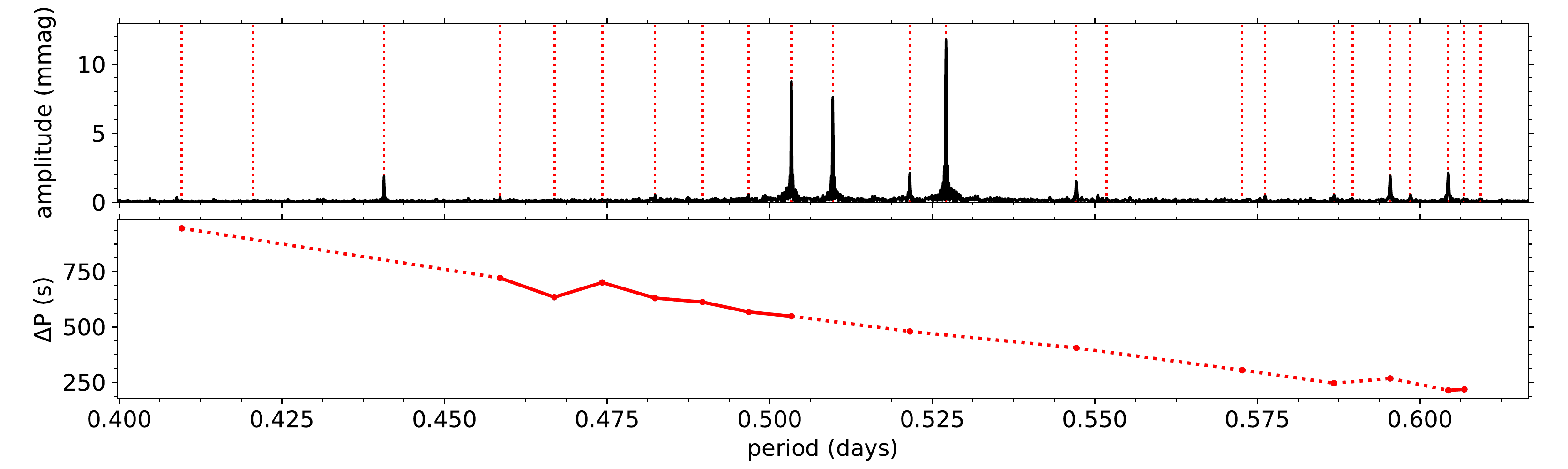}
    \caption{Detected period-spacing patterns of KIC\,12785282. {\em Top:} Lomb-Scargle periodogram (black) with the g~modes (red dotted lines) that form a $(k,m) = (0,1)$~pattern. {\em Bottom:} Period spacing as a function of pulsation period for the detected g-mode pattern, with dotted lines indicating gaps.\label{fig:kic12785282_period-spacings}}
    \includegraphics[width=\textwidth]{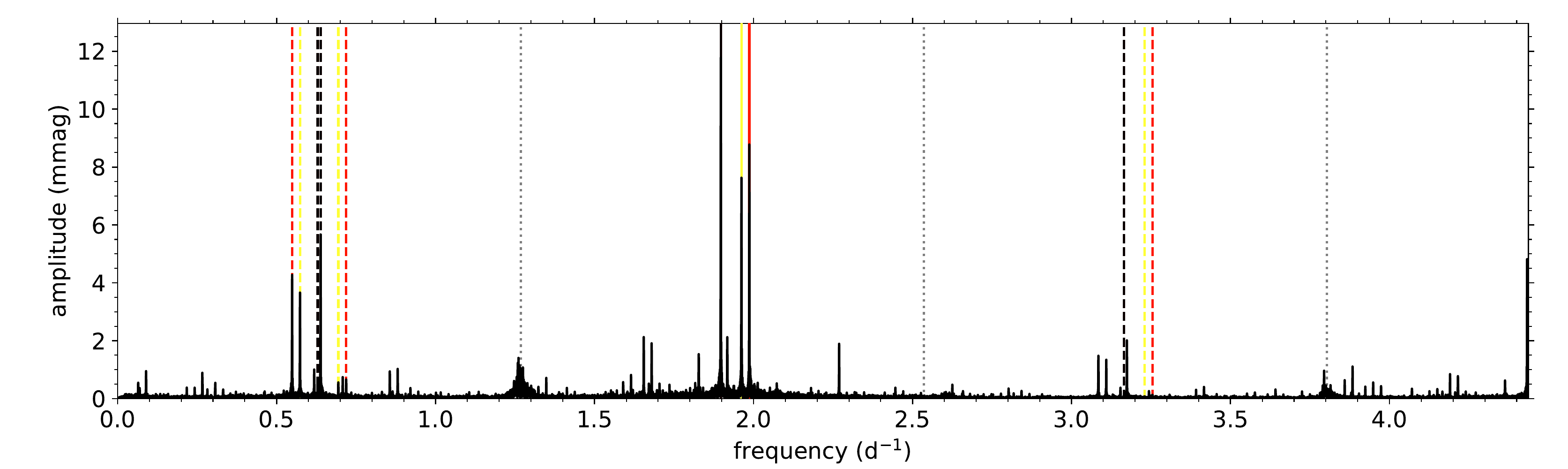}
    \caption{Orbital-frequency-spaced multiplets of KIC\,12785282. Lomb-Scargle periodogram (black) with the independent g-mode pulsations (full lines) and their multiplet components (dashed lines) shown in different matching colours. The grey dotted lines indicate orbital harmonics. \label{fig:kic12785282_fourier}}
\end{figure*}

\begin{figure}
    \centering
    \includegraphics[width=\columnwidth]{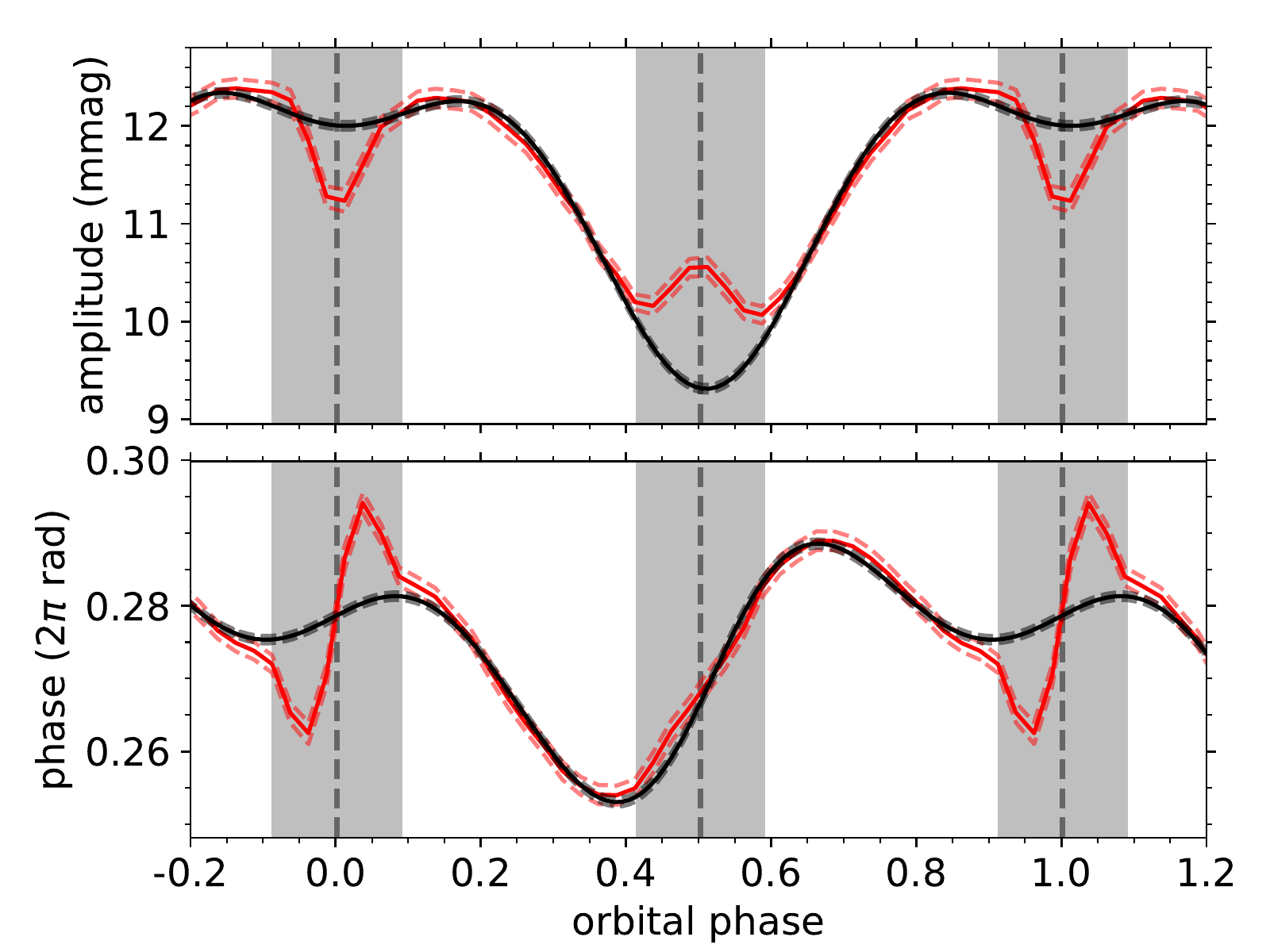}
    \caption{Tidal perturbation of the dominant $(k,m) = (0,1)$ g~mode of KIC\,12785282, with $\nu = 1.8971606(7)\,\rm d^{-1}$, similar to Fig.\,\ref{fig:kic03228863_bestpuls}.\label{fig:kic12785282_bestpuls}}
\end{figure}
\clearpage
\newpage


\section{Distorted g-mode amplitudes}
\label{appendix:toymodel}
To evaluate the properties of g-mode pulsations with tidally distorted pulsation amplitudes, we built a toy model\footnote{A python implementation of this algorithm is available at \href{https://github.com/TVanReeth/tidal-perturbation-simulations}{https://github.com/TVanReeth/tidal-perturbation-simulations}.} of a g-mode pulsator in a close binary, and artificially scaled the pulsation amplitude on the stellar surface. To simplify the calculations, it is assumed that the stars are spherically symmetric and rigidly rotating, and that the binary system is circularised and synchronised. Moreover, it is also assumed that the stellar rotation axis is perpendicular to the orbital plane. 

In the corotating frame of the pulsating star, we then define our reference axes such that the z-axis coincides with the rotation axis of the star, and the y-axis with the tidal axis of the binary. The local brightness variations $B\left(\theta,\phi,t\right)$ caused by the g-mode pulsation are then described by
\begin{equation}
    B\left(\theta,\phi,t\right) = A S_c\left(\theta,\phi\right) \Theta_{km}\left(\theta,s\right) \cos\left[2\pi\nu_{\rm co}t - m\phi\right],
\end{equation}
where $A$ is the amplitude of the pulsation relative to the (local) brightness of the star, and $S_c\left(\theta,\phi\right)$ is the distortion of the pulsation amplitude on the stellar surface. $\Theta_{km}\left(\theta,s\right)$ is the Hough function for the g~mode with identification $(k,m)$, calculated within the theoretical framework of the TAR, and the spin parameter $s = 2\nu_{\rm rot}/\nu_{\rm co}$ with $\nu_{\rm rot}$ and $\nu_{\rm co}$ being the rotation frequency and the pulsation frequency in the corotating frame, respectively. $(\theta,\phi)$ are the angular spherical coordinates with respect to the rotation axis. 

In our toy model, the local pulsation amplitudes on the stellar surface are distorted by the function
\begin{equation}
    S_c\left(\theta,\phi\right) = 1 + A'\sin\theta\sin\phi = 1 + A'\cos\theta_T,\label{eq:distortion}
\end{equation}
where $A'$ is the amplitude of the distortion and $\theta_T$ is the angle between the normal of the evaluated surface patch and the tidal axis.

Subsequently, we convert from the corotating to the inertial reference frame by defining a new set of spherical coordinates $(\theta_L,\phi_L)$ with respect to the observer's line-of-sight at time stamp $t$. This is done by replacing the azimuthal angle $\phi$ in the equations above by $\phi - 2\pi\nu_{\rm rot}$, and rotating the system around the inclination angle $i$. The flux $F_j(t)$ that the observer receives from the $j^{th}$ stellar component in the binary at time $t$, can then be calculated as  
\begin{equation}
   \begin{split}
    F_j\left(t\right) = \frac{1}{F_{\rm norm}}\int_{0}^{2\pi}\mathrm{d}\phi_L&\int_0^{\frac{\pi}{2}}\mathrm{d}\theta_L \sin\left(2\theta_L\right) \mu\left(1 - \cos\theta_L\right)\\ 
     & \times\left( 1 + B\left(\theta,\phi,t\right)\right) M_{\rm ecl}\left(\theta_L,\phi_L,t\right),
   \end{split}\label{eq:toypuls_int}
\end{equation}
where we have included linear limb darkening (with coefficient $\mu$). The mask $M_{\rm ecl}\left(\theta_L,\phi_L,t\right)$ indicates which parts of the stellar surface are not eclipsed by the other binary component, that is, which parts are visible to the observer at time $t$. Given the masses and radii of the binary components, the inclination angle $i$ of the system, and our assumption of spherically symmetric stars in a circular binary with synchronised rotation, these are straightforward to compute. Finally, the normalisation constant $F_{\rm norm}$ ensures that $ F_j\left(t\right) = 1$ when the star is non-pulsating ($A = 0$) and not eclipsed.

The observed flux variability of the pulsating binary system is then calculated as
\begin{equation}
    F\left(t\right) = L_1\cdot F_1\left(t\right) + L_2\cdot F_2\left(t\right) + L_3,
\end{equation}
where $F_1\left(t\right)$ and $F_2\left(t\right)$ indicate the normalised flux of the individual binary components, $L_1$ and $L_2$ are the relative light contributions from both components, and $L_3$ is the third light.

\begin{figure}
    \centering
    \includegraphics[width=\columnwidth]{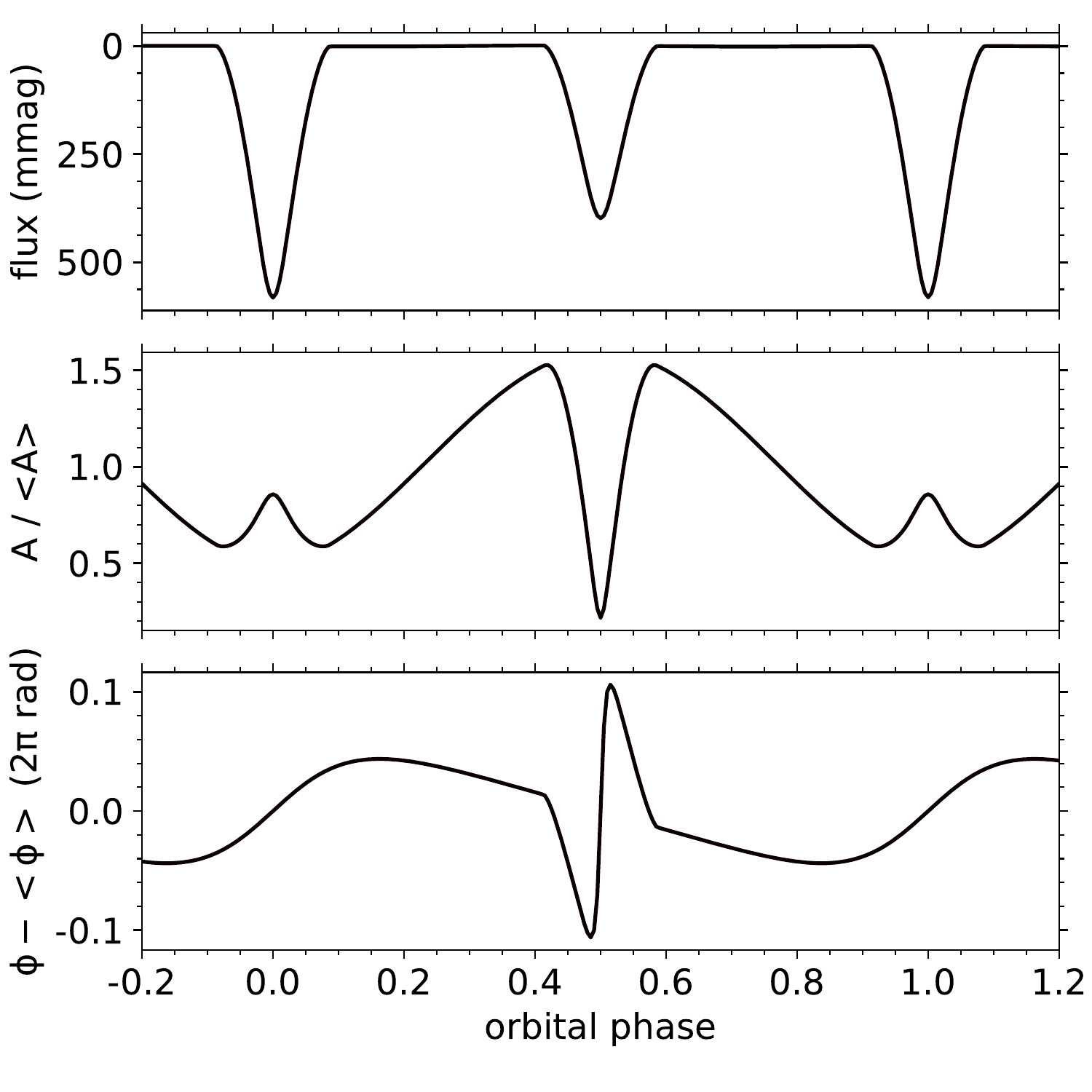}
    \caption{\label{fig:V456Cyg_model}High-resolution toy model of a $(k,m) = (0,1)$ g-mode pulsation with a distorted amplitude in the secondary component of a close binary system, based on the parameter values of V456\,Cyg, as listed in Table\,\ref{tab:V456Cyg_model}. {\em Top:} the flux variability caused by the binary motion. Both stellar components are assumed to be spherical, leading to the absence of ellipsoidal variability in this model. {\em Middle:} the observed amplitude modulation as a function of the orbital phase. {\em Bottom:} the observed pulsation phase modulation as a function of the orbital phase, caused by the geometry of the pulsation amplitude distortion across the stellar surface.}
\end{figure}

To illustrate our toy model, we have built a simulation using stellar parameter values that approximate those measured for V456\,Cyg \citep[listed in Table \ref{tab:V456Cyg_model}; obtained from][]{VanReeth2022a}. A $(k,m) = (0,1)$ g-mode pulsation has been included for the secondary component, with the amplitude distorted as described by Eq.(\ref{eq:distortion}) with $A' = 0.667$, so that the pulsation is up to five times larger in the stellar hemisphere facing the primary than on the opposite side. The simulated light curve was subsequently analysed using the same methodology as described in Section\,\ref{sec:methods}. The result, shown in high resolution in Fig.\,\ref{fig:V456Cyg_model}, is consistent with the observed tidal g-mode perturbations of V456\,Cyg, as presented in Fig.\,5 of \citet{VanReeth2022a}. Most notably, the amplitude modulations during the out-of-eclipse phases are reflected in the pulsation phase modulations, with the phase decreasing during the orbital phases when the pulsation amplitude is maximal, as observed. Similarly, the signatures of spatial filtering are clearly reproduced during the eclipses.

A posteriori, we can also see that the assumption of spherical symmetry in our toy model was justified. Real tidally distorted stars in close binaries like V456\,Cyg exhibit ellipsoidal and rotational-modulation-like variability, caused by tidal distortion of stars in close binaries. The associated stellar flux variations also modulate the apparent pulsation mode amplitude on the stellar surface, similar to the local pulsation amplitude distortion that we simulated in our toy model. However, as ellipsoidal variability is usually small (of the order of 10 to 100\,mmag for the stars in this work), the associated pulsation amplitude modulations will also be much smaller than what we see in our observations or what was simulated in our model.

\begin{table}[ht!]
	\centering
	\caption{Toy model parameter values used in the simulation shown in Fig.\ref{fig:V456Cyg_model}, based on the parameters values of V456\,Cyg \citep{VanReeth2022a}.}
	\label{tab:V456Cyg_model}
	\begin{tabular}{ll} 
		\hline\hline\vspace{-9pt}\\
		   Parameter       &       Value \\
		\hline\vspace{-9pt}\\
		orbital frequency $\nu_{\rm orb}$ ($\rm d^{-1}$) & 1.122091 \\
		inclination angle $i$ (${}^\circ$) & 83.198\\
		stellar mass $M_1$ ($M_\odot$) & 1.859\\
		stellar mass $M_2$ ($M_\odot$) & 1.576\\
		stellar radius $R_1$ ($R_\odot$) & 1.588\\
		stellar radius $R_2$ ($R_\odot$) & 1.507\\
		linear LD coefficient $\mu_1$ & 0.248\\
		linear LD coefficient $\mu_2$ & 0.285\\
        light contribution $L_1$ & 0.6 \\
        light contribution $L_2$ & 0.4 \\
        third light $L_3$ & 0.0 \\
        \hline\vspace{-9pt}\\
        pulsation frequency $\nu_{\rm puls,in}$ ($\rm d^{-1}$) & 1.9491 \\
        pulsation amplitude $A$ & 0.005\\
        amplitude distortion $A'$ & 0.667\\
        mode identification $(k,m)$ & (0,1)\\
		\hline
	\end{tabular}
    \tablefoot{The binary parameters are listed in the top half of the table, the asteroseismic ones in the bottom half.}
\end{table}

\end{appendix}

\end{document}